\documentclass[10pt, superscriptaddress, aps, pra, twocolumn, longbibliography, showkeys]{revtex4-2}
\usepackage{braket}
\usepackage{amsmath,amssymb,amsthm}
\usepackage{easybmat}
\usepackage{tabularx} 
\usepackage{array}
\usepackage[pdftex]{graphicx}
\usepackage{graphicx}
\usepackage{float}
\usepackage{placeins}
\usepackage{diagbox} 
\usepackage{times,txfonts}
\usepackage{color}
\usepackage{xcolor}
\usepackage{appendix}
\usepackage{natbib}
\usepackage[colorlinks=true, citecolor=blue, urlcolor=blue, linkcolor=blue]{hyperref}

\begin{document}

\title{Two-Qubit Spin-Boson Model in the Strong Coupling Regime: Coherence, Non-Markovianity, and Quantum Thermodynamics}
\author{Hasan Mehdi Rizvi\textsuperscript{}}
\email{hasanmehdi.1102@gmail.com }
	\author{Devvrat Tiwari\textsuperscript{}}
	\email{devvrat.1@iitj.ac.in }
    \thanks{H.M.R. and D.T. contributed equally to this work.}
    \author{Subhashish Banerjee\textsuperscript{}}
	\email{subhashish@iitj.ac.in }
	\affiliation{Indian Institute of Technology Jodhpur-342030, India\textsuperscript{}}

\date{\today}


\begin{abstract}
We investigate the dynamics of a two-qubit open quantum system, in particular the two-qubit spin-boson model in the strong coupling regime, coupled to two thermal bosonic baths under non-Markovian and non-equilibrium conditions. Two complementary approaches, the Hierarchical Equations of Motion (HEOM) and Reaction Coordinate Mapping (RCM), are employed to examine various coupling regimes between the qubits and their respective baths. The dynamical features of the model and the impact of the tunneling amplitude on quantum coherence of the system are probed using the $l_1$-norm of coherence. The model is further shown to have non-Markovian evolution. The nontrivial task of calculating entropy production in the strong-coupling regime is performed using auxiliary density operators in HEOM. Motivated by the realization of a quantum thermal device in the strong-coupling regime, the non-equilibrium steady-state behavior of the system is investigated. Furthermore, the relationship between the heat and spin currents and the tunneling amplitude is probed. 
\end{abstract}
\maketitle

\section{Introduction}\label{intro}

The theory of open quantum systems deals with the dynamics of a quantum system interacting with its ambient environment~\cite{Breuer2002, sbbook, weiss}. These systems find use in a number of applications~\cite{CaldeiraLeggett1981, Louisell, GrabertSchrammIngold, sbsterngerlach, sbqbm, sbdixit_2019, sbthomas, tiwari2025_finite_bath}. A traditional route to studying these systems has been to use the Gorini-Kossakowski-Sudarshan-Lindblad master equation, modeling a Markovian dynamics~\cite{gksl, lindblad}. However, recently, interest in the challenging domain of non-Markovian dynamics has surged~\cite{de_Vega_2017, Rivas_2014, hcla_non_Markovianity, colloquium_NM_breuer, Kumar_2018, Utagi_2020, kading2025, tiwari_strong}. The spin-boson model, describing the dynamics of a two-level system interacting with a bosonic bath, has been central to these studies~\cite{weiss, Leggett1987, THORWART2004}. It is a paradigmatic non-integrable model that describes the dissipative and decoherence dynamics of an open quantum system and has been experimentally realized in various contexts~\cite{Sun2025, Mostame_2012, spin_boson_expt2, Magazzu2018}. Determining the dynamics of the system in this model is highly nontrivial, particularly in the strong coupling regime. In the weak-coupling regime under the secular approximation, the spin-boson model has been shown to converge to a phase-covariant dynamics that remains non-Markovian~\cite{Haase2018, Goan_2010, Goan_2011, mahima2025}. To obtain the dynamics of a spin-boson system in the strong coupling regime, several numerical techniques have been developed over the years, including reaction-coordinate mapping (RCM)~\cite{IlesSmith2014, Strasberg2016, Dvira_rcm, Shubrook_2025}, hierarchical equation of motion (HEOM)~\cite{tanimura-kubo, Tanimura_exact_2020, tanimura_two_qubit_paper, tanimura_heat_current, HierarchicalEOM, efficient_heom}, the path integral approach~\cite{weiss, Sassetti_1990, Sassetti_1996, Amartya_bose_2023}, polaron transformation~\cite{Chen2020_polaron}, non-equilibrium Green’s functions~\cite{Esposito2015_NEGF}, and the pseudomode technique~\cite{Garraway_pseudomode1, Linden_pseudomode2, Plenio_pseudomode3}, among others.  

At the intersection of quantum information theory and the dissipative dynamics of open quantum systems, the traditional laws of thermodynamics are reformulated, giving rise to the field of quantum thermodynamics~\cite{Gemmer2009, kosloff, deffner_book, Landi2021}. Particularly for strongly coupled open quantum systems, correctly defining fundamental quantum thermodynamic quantities, such as entropy production, heat, and work, is an active area of research~\cite{tiwari_strong, Strasberg2016, Rivas_strong_coupling, Landi2021}. The non-equilibrium behavior in quantum thermodynamics becomes essential for understanding energy exchange in quantum thermal devices at the nanoscale, and the subject of non-equilibrium thermodynamics is driven by the physical realization of these devices~\cite{Landi_ness_paper, Baowen_review, Dubi_review, BENENTI2017}.

Two qubits, each coupled to its own thermal bosonic bath maintained at different temperatures, form a minimal yet versatile non-equilibrium setup for probing open-system dynamics~\cite{Landi_ness_paper}. These systems often reach a non-equilibrium steady state (NESS), characterized by nonzero thermal transport between the baths. Quantum thermal devices, such as thermal diodes, transistors, Wheatstone bridges, and thermal adders, arise from thermal transport in a NESS~\cite{Segal_spin_boson_rectifier, diode1, thermal_diode, transistor, QWB, Tiwari2025_circuit}, albeit in a weak coupling and Markovian setting. Due to advancements in technology, these boundary-driven systems are now directly accessible in a laboratory~\cite{Landi_ness_paper, Majland_2020, Cerisola_2025}. In this work, we consider a two-qubit spin-boson model setup, as shown in Fig.~\ref{fig_Schematic_two_qubit_spin_boson_model}, motivated by the realization of a quantum thermal device in the strong-coupling regime. To obtain the dynamics of this two-qubit spin boson model, we employ the RCM and the HEOM techniques. The RCM is used to extract a dominant collective bath mode, the reaction coordinate, exactly; the residual environment is then treated perturbatively via a second-order master equation. This hybrid strategy retains the mapping’s exact treatment of the primary bath coupling while offering analytic tractability for the remainder. The HEOM technique unravels the bath's memory effects by introducing a hierarchy of coupled auxiliary density operators, thereby transforming the non-Markovian dynamics into a numerically solvable system of local differential equations. This non-perturbative framework captures arbitrary system-bath coupling strengths and memory timescales. In this work, first a comparison between the two techniques is made, and, due to the numerical convenience of the HEOM technique, it is used throughout the rest of the work. 

With the dynamics of the system under control, we investigate the interesting dynamical features of the model, including coherence (using $l_1$-norm of coherence~\cite{Baumgratz_2014}) and non-Markovian evolution (using the BLP measure~\cite{BLP_2009}). Before launching investigations into the NESS behavior of the system, we study key quantum thermodynamic quantities, including entropy and entropy production, and establish the validity of the second law of quantum thermodynamics. To probe the NESS behavior of the system, we examine the non-zero steady-state coherence, and we inspect thermal transport between the baths using spin and heat currents. These currents are the key quantities in a quantum thermal circuit. Recently, it has been demonstrated that they follow quantum thermal versions of Kirchhoff's current laws in a quantum thermal circuit~\cite{Tiwari2025_circuit}. Furthermore, they are generally related by a scaling factor. In this work, this relationship is examined in the presence of a nonzero tunneling amplitude.  

The article is structured as follows. In Sec.~\ref{setup}, the two-qubit spin boson model setup is discussed along with various coupling regimes, and a comparison is made between the RCM and the HEOM techniques. Quantum coherence and non-Markovian evolution of the system are explored in Sec.~\ref{sec_coherence_non_Markovian}. Section~\ref{sec_q_thermo} discusses the entropy, entropy production, second law of thermodynamics, and NESS, including spin and heat currents. The conclusions are presented in Sec.~\ref{sec_conclusions}. 

\section{Two-qubit spin-boson model}\label{setup}
\begin{figure}
  \centering
  \includegraphics[width=0.9\linewidth]{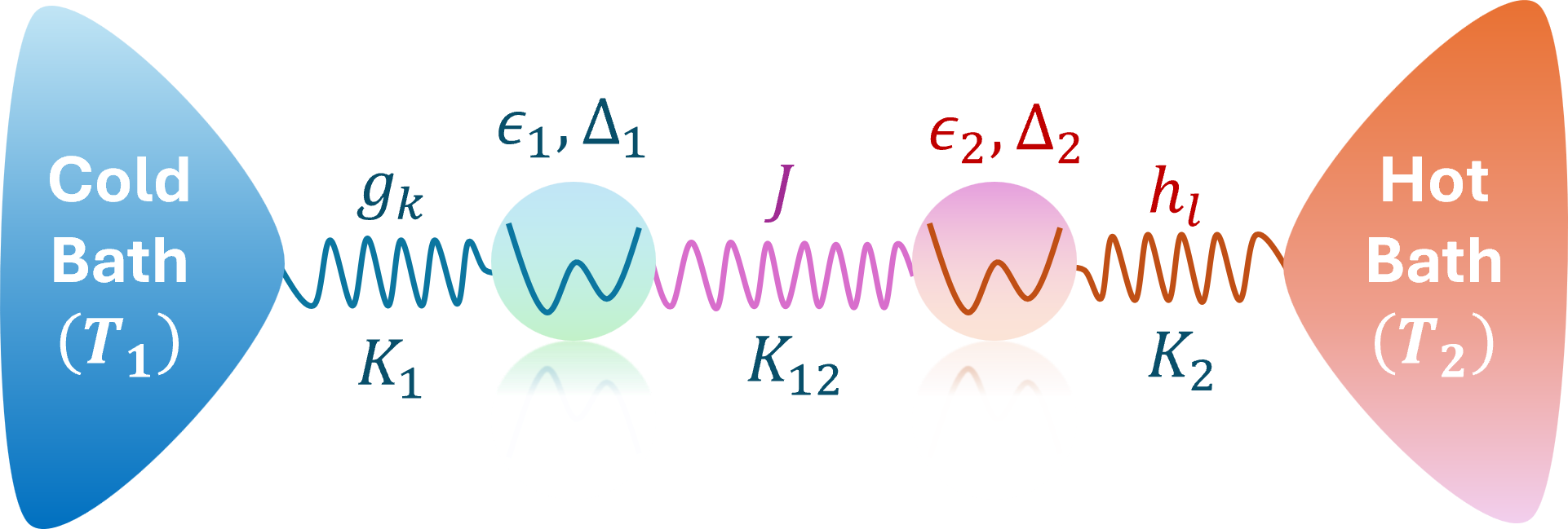}  
  \caption{A schematic diagram of the two-qubit spin-boson model, depicting an interaction of the qubits with their respective bosonic baths at different temperatures. $K_1, K_{12}$, and $K_2$ characterize the interaction strengths between qubit 1 and the cold bath, between qubits 1 and 2, and between qubit 2 and the hot bath, respectively.}
  \label{fig_Schematic_two_qubit_spin_boson_model}
\end{figure}

Here, we briefly discuss the two-qubit spin-boson model (see Fig.~\ref{fig_Schematic_two_qubit_spin_boson_model}) and the corresponding schemes, viz., the reaction-coordinate mapping (RCM) and hierarchical equation of motion (HEOM) approaches, for solving it in the strong-coupling regime. 
The total Hamiltonian of the two-qubit spin-boson model is given by
\begin{align}\label{eq_total_system_Hamiltonian}
    H &= H_S + H_B + H_{SB},
\end{align}
where the system Hamiltonian $H_S$ is given by
\begin{align}
    H_S &= H_1 + H_2 + H_{12}\nonumber \\
    &= \sum_{i = 1}^{2}\left(\frac{\epsilon_{i}}{2}\sigma_{z}^{(i)} + \frac{\Delta_{i}}{2}\sigma_{x}^{(i)}\right) 
    + J\left(\sigma_{x}^{(1)}\sigma_{x}^{(2)} + \sigma_{y}^{(1)} \sigma_{y}^{(2)}\right),
\end{align}
the bath Hamiltonian $H_B$ is given by
\begin{align}
     H_B &= H_{B_1} + H_{B_2}\nonumber \\
     &=\sum_{k}\left[\frac{1}{2}\left(p_{k}^{2} + \omega_{k}^{2}q_{k}^{2}\right)\right] 
    + \sum_{l}\left[\frac{1}{2}\left(\pi_{l}^{2} + \gamma_{l}^{2}x_{l}^{2}\right)\right],
\end{align}
and the system-bath interaction Hamiltonian $H_{SB}$ is given by
\begin{align}
    H_{SB} &= H_{SB_1} + H_{SB_2} = \sigma_{z}^{(1)}\sum_{k}g_{k}q_{k} + \sigma_{z}^{(2)}\sum_{l}h_{l}x_{l}.
\end{align}
Here, \( \epsilon_i \) is the energy splitting of the \(i\)-th qubit, \( \Delta_i \) is the tunneling amplitude (transverse field strength) of the \(i\)-th qubit, \( \sigma_{x}^{(i)}, \sigma_{y}^{(i)}, \sigma_{z}^{(i)} \) are the Pauli matrices corresponding to the \(i\)-th qubit, \( J \) is the inter-qubit coupling strength (for Heisenberg $XX$ type interaction), \( p_k, q_k \) are the canonical momentum and position operators of the \(k\)-th oscillator in the bath coupled to qubit 1, and \( \omega_k \) is the frequency of the \(k\)-th mode in the bath coupled to qubit 1. Furthermore, \( \pi_l, x_l \) are the canonical momentum and position operators of the \(l\)-th oscillator in the bath coupled to qubit 2, and \( \gamma_l \) is the frequency of the \(l\)-th mode in the bath coupled to qubit 2. In the interaction Hamiltonian $H_{SB}$, \( g_k \) is the coupling strength between the \(k\)-th oscillator and qubit 1, and \( h_l \) is the coupling strength between the \(l\)-th oscillator and qubit 2. The baths connected to the qubits are maintained at different temperatures, as shown in Fig.~\ref{fig_Schematic_two_qubit_spin_boson_model}. $T_{1(2)}$ is the temperature of the bath connected to qubit 1(2). 

In Fig.~\ref{fig_Schematic_two_qubit_spin_boson_model}, $K_1$, $K_{12}$, and $K_{2}$ describe the interaction strength between qubit 1 and the cold bath, between qubits 1 and 2, and between qubit 2 and the hot bath. Thus, for example, $K_1=$ W, $K_{12}=$ S, $K_2=$ W (acronyms W and S stand for weak and strong couplings, respectively) corresponds to the scenario when the qubits are weakly coupled to their respective baths, but the interaction between the qubits is strong. Here, we shall investigate four regimes of couplings, which are $K_1~K_{12}~K_2 = \{\text{WWW, WSW, SWS, SSS}\}$. Furthermore, we provide the parameters applicable to the above four regimes, and used throughout the paper, unless stated otherwise. In the WWW regime, \(T_{1} = T_{\rm cold} = 1.04\),\,\(T_{2} = T_{\rm hot} = 1.39\), \(\Delta_1 = 2.0\), \(\epsilon_1 = 1.0\), \(\omega_{c_1} = 0.05\Delta_1\), \(\Delta_2 = 1.6\), \(\epsilon_2 = 0.75\), \(\omega_{c_2} = 0.10\Delta_2\) and \(\alpha_{1} = \alpha_{2} = J = 0.05/\pi\), where \( \alpha_{1} \) and \( \alpha_{2} \) denote the coupling strengths of qubits 1 and 2 with their respective thermal baths and $\omega_{c_i}$'s are the respective cut-off frequencies, see Appendix A. In the SWS regime, $\alpha_1 = \alpha_2 = 2.5/\pi, J = 0.05/\pi$, and the rest of the parameters remain the same. Similarly, in the WSW regime, $\alpha_1 = \alpha_2 = 0.05/\pi$, and $J = 2.5/\pi$ and in the SSS regime, $\alpha_1 = \alpha_2 = J = 2.5/\pi$.  

Two techniques, the RCM and the HEOM techniques, are used here to solve this two-qubit spin-boson model. In the RCM technique, an RC-mapped master equation is obtained in Appendix A and which governs the dynamics of the model. The {\it HierachicalEOM.jl} package~\cite{HierarchicalEOM}, based on the {\it QuantumToolbox.jl} package~\cite{QuantumToolbox} in Julia, is used to simulate the exact dynamics of the two-qubit spin-boson model using the HEOM technique. The HEOM technique is briefly sketched in Appendix B.

\begin{figure}
    \centering
    \includegraphics[width=1\linewidth]{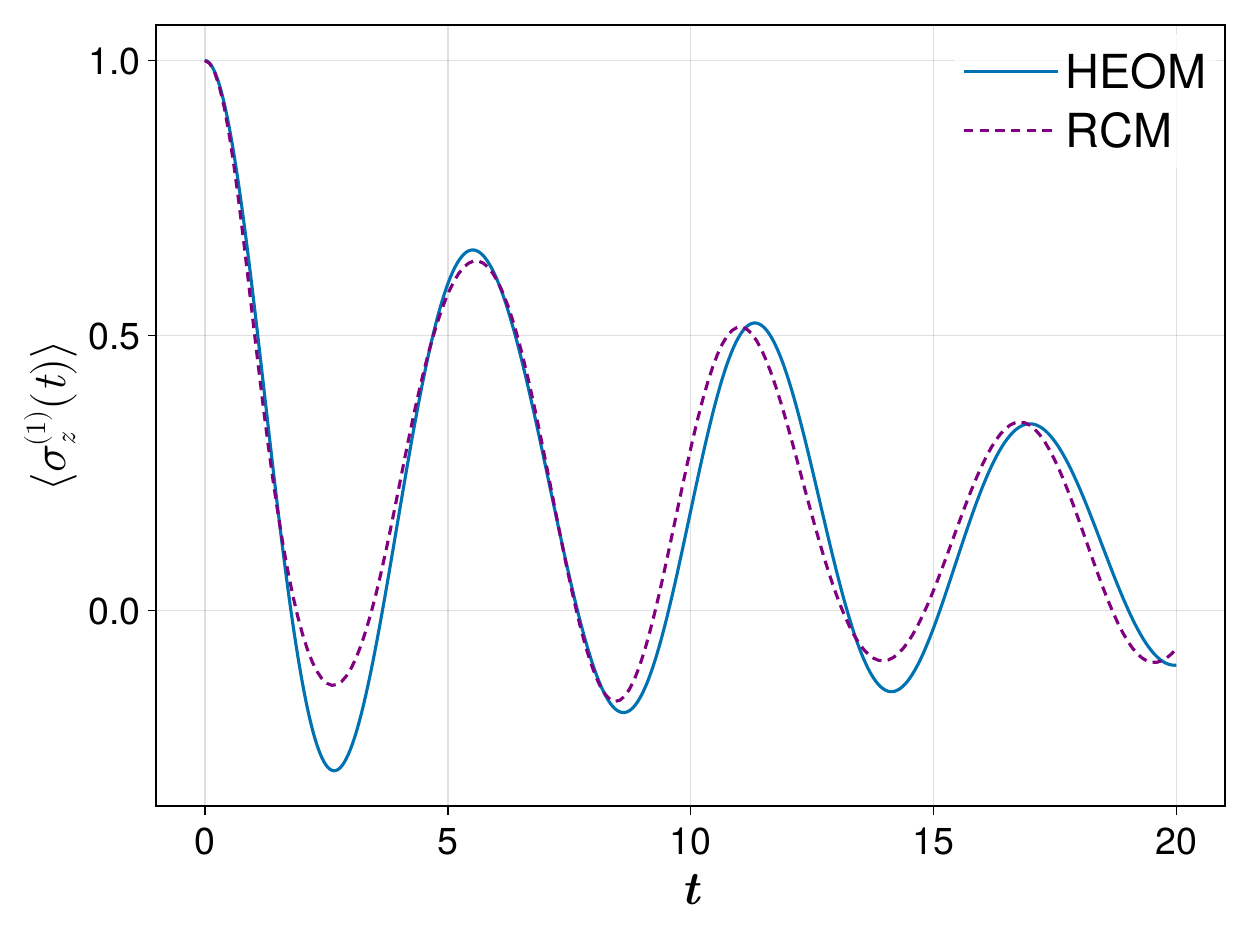}
    \caption{Variation of the expectation value $\left\langle\sigma^{(1)}_z\right\rangle$ with time using both RCM and HEOM techniques. The parameters are taken to be: $\Delta_1 = 1.0, \epsilon_1 = 0.5\Delta_1, \Delta_2 = 1.0,\epsilon_2 = 0.4\Delta_2, J = 0.1/\pi, \alpha_1 = 0.2/\pi, \alpha_2 = 0.2/\pi, \omega_{c_1} = 0.05\Delta_1, \omega_{c_2} = 0.10\Delta_2, T_1 = 1.04, T_2 = 1.39$. The excited state is taken to be the initial state of the system.}
    \label{fig_comparison_rcm_heom}
\end{figure}
Figure~\ref{fig_comparison_rcm_heom} shows a comparison between the RCM and the HEOM techniques, where the variation of the expectation value $\left\langle\sigma^{(1)}_z\right\rangle = {\rm Tr}\left[\rho_S(t)\left(\sigma_z^{(1)}\otimes\mathbb{I}^{(2)}\right)\right]$ with time is plotted using both methods. Both the RCM and the HEOM techniques show significantly similar variation. Here, we note that the RCM technique requires more numerical computation time than the HEOM technique when solving the two-qubit spin-boson model with similar accuracy. Due to this, in the subsequent part of the paper, we employ the HEOM technique to study coherence, non-Markovianity, and quantum thermodynamics of the two-qubit spin-boson model.   
\section{Quantum coherence and non-Markovian evolution}\label{sec_coherence_non_Markovian}
Here, we study the quantum coherence and non-Markovianity of the two-qubit spin-boson model in the WWW, SWS, WSW, and SSS regimes (outlined in Sec.~\ref{setup}). To study coherence, we employ the $l_1$-norm~\cite{Streltsov_2017}. The trace distance measure, also known as the Breuer-Laine-Piilo (BLP) measure~\cite{BLP_2009}, is used as a witness of non-Markovian evolution. 

\subsection{Coherence}
Quantum coherence is a fundamental property of quantum systems that distinguishes quantum physics from classical physics. The $l_1$-norm of coherence $C_{l_1}$ is one of the prominent quantum coherence quantifiers, and is given by~\cite{Baumgratz_2014} 
\begin{align}\label{eq_l1_norm}
    C_{l_1} = \sum_{i\ne j} |\rho_{ij}|,
\end{align}
where $\rho_{ij}$ are the off-diagonal elements of the system's density matrix. Here, the HEOM method (see Sec.~\ref{setup}) is used to obtain the reduced density operator of the two-qubit spin-boson model, and to calculate the $l_1$-norm of coherence $C_{l_1}$.
\begin{figure}
    \centering
    \includegraphics[width=1\linewidth]{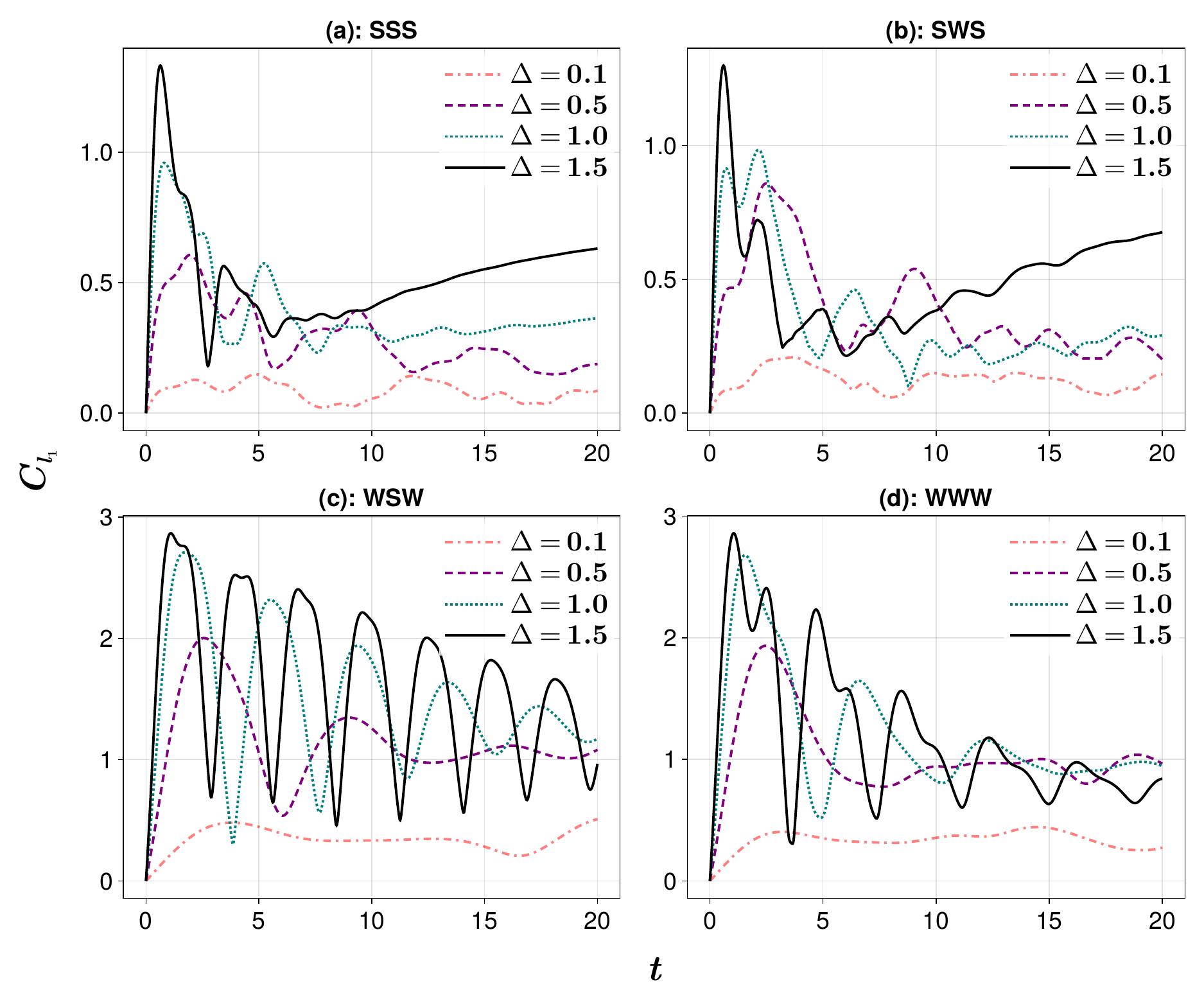}
    \caption{Variation of $l_1$-norm of coherence ($C_{l_1}$) with time for different values of tunneling amplitudes $\Delta_1 = \Delta_2 = \Delta$ and in (a) SSS, (b) SWS, (c) WSW, and (d) WWW coupling regimes. $\ket{00}_{12}$, where $\ket{0} = \begin{pmatrix}
        1\\0
    \end{pmatrix}$, is taken to be the initial state of the two-qubit system. The parameters are discussed in Sec.~\ref{setup}.}
    \label{fig_tqsbm_coherence}
\end{figure}

Figure~\ref{fig_tqsbm_coherence} depicts the variation of $C_{l_1}$ with time for different values of the tunneling amplitudes $\Delta = \Delta_1 = \Delta_2$ in the SSS, SWS, WSW, and WWW coupling regimes. To study the emergence of coherence, a coherence-less $\ket{00}_{12}$ state, where $\ket{0} = \begin{pmatrix}
    1\\0
\end{pmatrix}$, is taken as the initial state of the system. The figure shows that coherence emerges across all coupling regimes. A common feature across regimes is that coherence reaches its largest values at the highest $\Delta$ and its smallest values at the lowest $\Delta$. In the SSS and SWS regimes, a disordered oscillatory behavior emerges, which can be attributed to pronounced decoherence resulting from stronger coupling to the bath, see Figs.~\ref{fig_tqsbm_coherence}(a) and (b). In both these regimes, at later times, $C_{l_1}$ achieves an order in tandem with the value of $\Delta$. Interestingly, in the WSW and WWW regimes, Figs.~\ref{fig_tqsbm_coherence}(c) and (d), an ordered oscillation in the coherence is observed, which is prominently apparent when the inter-qubit coupling is strong, Fig.~\ref{fig_tqsbm_coherence}(c). Furthermore, in both WSW and WWW regimes, coherence generally shows a higher initial rise than in SSS and SWS regimes, due to the bath's lesser impact on the system's evolution. Furthermore, the WSW regime, where baths are weakly coupled, and the inter-qubit interaction strength is strong, emerges as the favorable regime for sustaining coherence.

\begin{figure}
    \centering
    \includegraphics[width=1\linewidth]{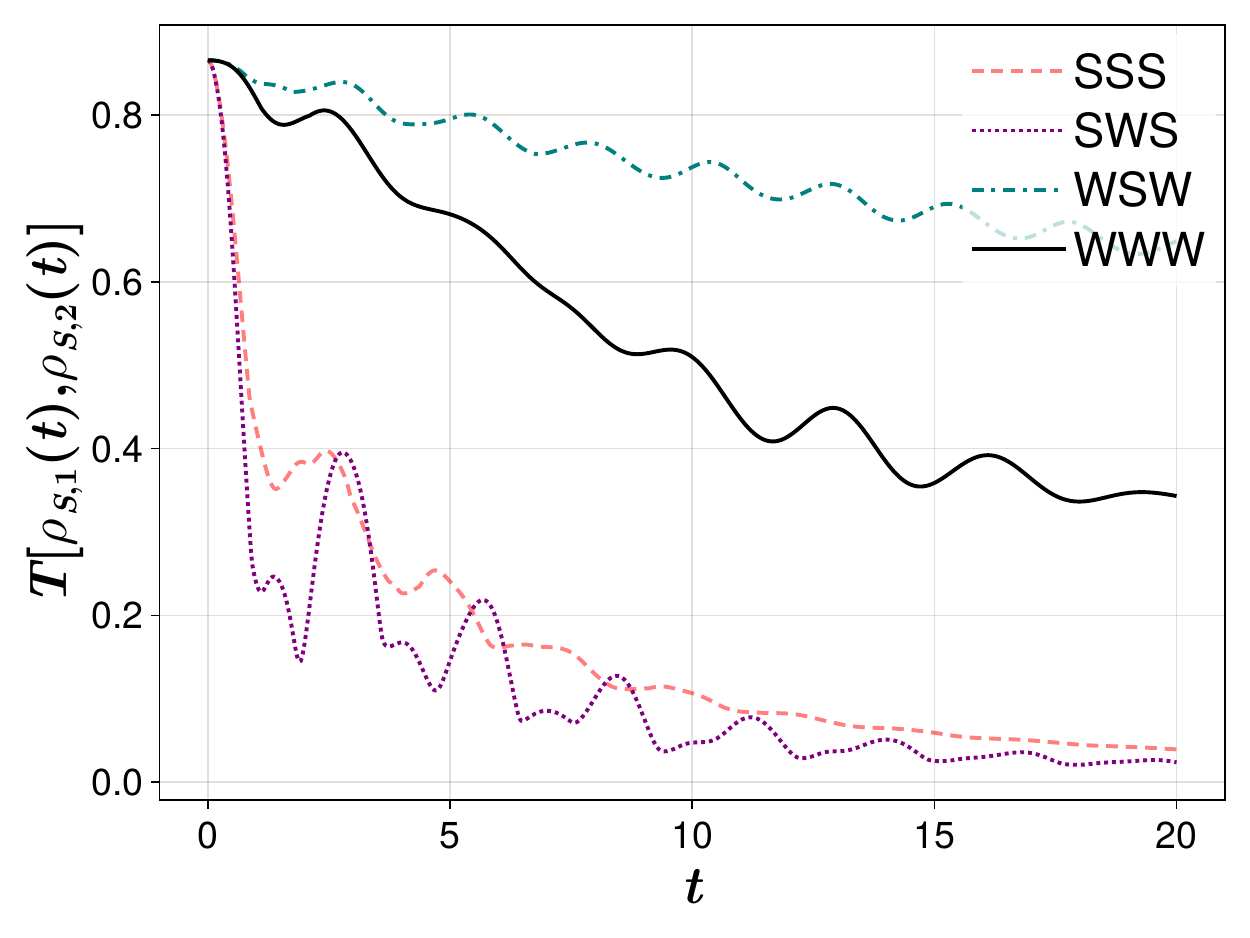}
    \caption{Variation of trace distance $T\left[\rho_{S, 1}(t), \rho_{S, 2}(t)\right]$ with time for different coupling regimes. Here, $\ket{00}_{12}$, where $\ket{0} = \begin{pmatrix}
        1\\0
    \end{pmatrix}$, and $\ket{++}_{12}$, where $\ket{+} = \frac{1}{\sqrt{2}}\begin{pmatrix}
        1\\1
    \end{pmatrix}$, are taken as the initial states of the two-qubit system. The parameters for different coupling regimes are discussed in Sec.~\ref{setup}.}
    \label{fig_blp_measure}
\end{figure}
\subsection{Non-Markovianity}
In general, an open quantum system exhibits non-Markovian dynamics~\cite{kading2025}. Several methods have been developed to quantify the non-Markovianity of an open quantum system~\cite{BLP_2009, Rivas_2014, Utagi_2020, de_Vega_2017}. Here, the trace distance measure (also known as the BLP measure) is used to investigate the non-Markovianity of the system's evolution~\cite{BLP_2009}. 

The trace distance is a measure of distinguishability between two quantum states, say $\rho_1$ and $\rho_2$, and is defined as 
\begin{align}
    T[\rho_1, \rho_2] = \frac{1}{2}\|\rho_1 - \rho_2\|_1 = \frac{1}{2}{\rm Tr}\left[\sqrt{\left(\rho_1 - \rho_2\right)^\dagger \left(\rho_1 - \rho_2\right)}\right],
\end{align}
where $\|(\cdot)\|_1 = {\rm Tr}[\sqrt{(\cdot)^\dagger(\cdot)}]$ is the trace norm of $(\cdot)$. A dynamical decrease in $T[\rho_1, \rho_2]$ can be interpreted as a loss of information from the open quantum system into the environment, which is a characteristic of Markovian dynamics. Revivals in the trace distance $T[\rho_2, \rho_2]$ indicate a flow of information from the environment back to the system, signifying non-Markovian behavior. In brief, the evolution of the system is non-Markovian if the trace distance exhibits revivals. 

Here, we consider two different initial states of the system, $\rho_{S, 1}(0) = \ket{00}_{12}$, where $\ket{0} = \begin{pmatrix}
    1\\0
\end{pmatrix}$ and $\rho_{S, 2}(0) = \ket{++}_{12}$, where $\ket{+} = \frac{1}{\sqrt{2}}\begin{pmatrix}
    1\\1
\end{pmatrix}$ and evolve them using the HEOM technique, outlined in Sec.~\ref{setup}. Their trace distance evolution $T\left[\rho_{S, 1}(t), \rho_{S, 2}(t)\right]$, in the SSS, SWS, WSW, and WWW coupling regimes, is plotted in Fig.~\ref{fig_blp_measure}. 
From the figure, it can be observed that in all regimes, the trace distance exhibits fluctuations, indicating a non-Markovian evolution. However, in the SWS regime, revivals in the trace distance are more pronounced, indicating stronger backflow of information. This shows that the strength of non-Markovianity is greater when the system is strongly coupled to the bath, but the inter-qubit coupling is weak. Furthermore, a general trend emerges: the trace distance decreases rapidly at early times when the system-bath coupling is strong, compared to the case of weak coupling. The trace distance maintains its maximum when the inter-qubit coupling is strong, and the system is weakly coupled to the bath, that is, in the WSW regime. 

\section{Quantum Thermodynamics}\label{sec_q_thermo}
A major objective of this work is to study the key thermodynamic features of the two-qubit spin-boson model in the arbitrary coupling regime. In this section, we delve deep into the study of von Neumann entropy and entropy production, the emergence of non-equilibrium steady states, and investigate the relationship between heat and spin currents. 

\subsection{Entropy and entropy production}
\begin{figure}
    \centering
    \includegraphics[width=1\linewidth]{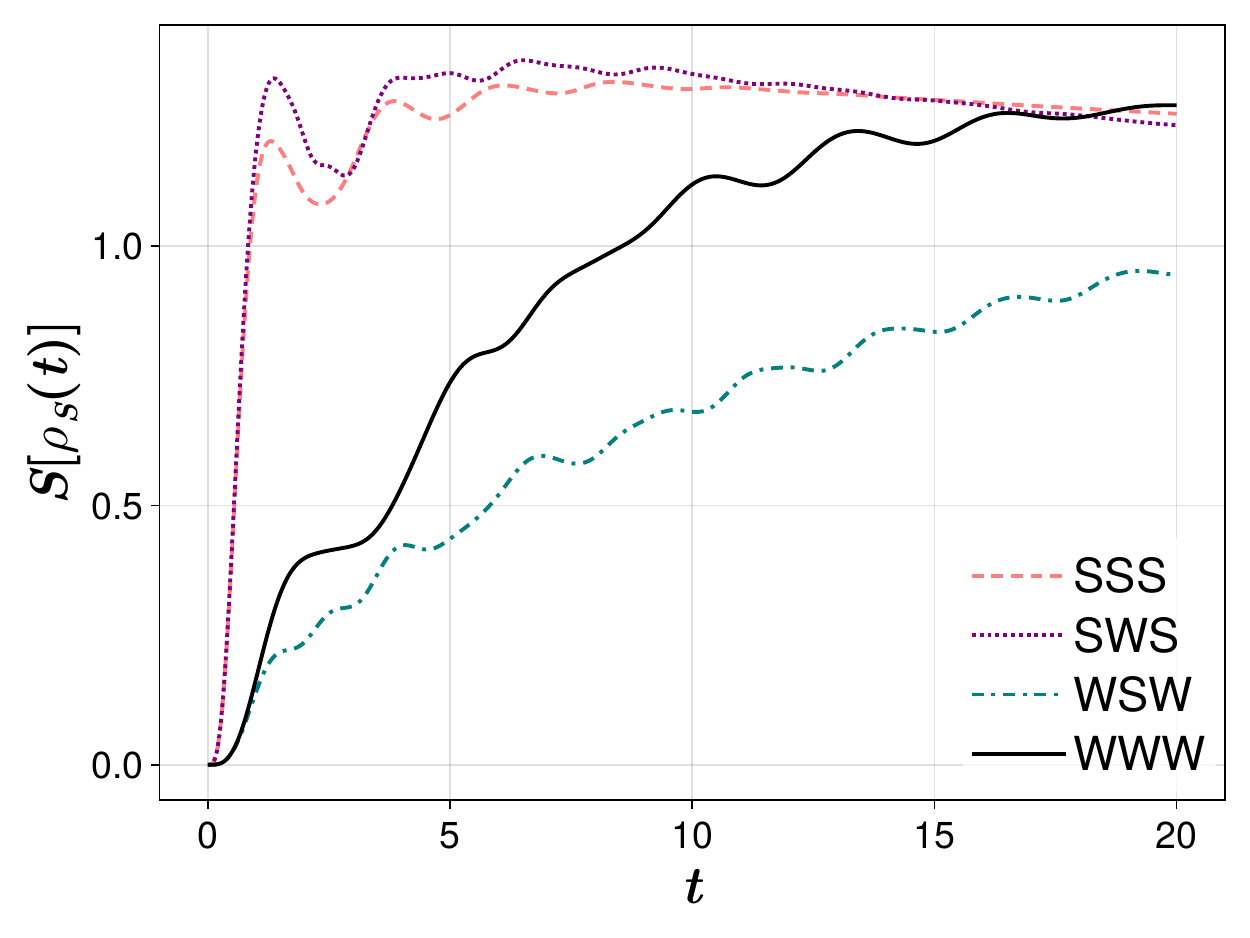}
    \caption{Variation of von Neumann entropy $S[\rho_S(t)]$ with time for the two-qubit spib-boson model in SSS, SWS, WSW, WWW coupling regimes. $\ket{00}_{12}$ is taken as the initial state of the system.}
    \label{fig_von_Neumann_entropy}
\end{figure}
\subsubsection{Entropy}
The von Neumann entropy, the quantum analogue of classical Shannon entropy, quantifies the information content or mixedness of a quantum state~\cite{Nielsen2010}, and is given by
\begin{align}\label{von_Neumann_entropy}
    S[\rho_S(t)] = -{\rm Tr}\left[\rho_S(t)\log\left\{\rho_S(t)\right\}\right].
\end{align}
The von Neumann entropy is zero for a pure quantum state and attains its maximum value for the maximally mixed state. Here, the von Neumann entropy is calculated from the state of the two-qubit system obtained via the HEOM technique, and its time evolution is studied. The logarithm of the state is computed directly using the {\it logm} function of the {\it QuantumToolbox.jl}~\cite{QuantumToolbox}. The variation of von Neumann entropy with time in the SSS, SWS, WSW, and WWW regimes is plotted in Fig.~\ref{fig_von_Neumann_entropy}. The von Neumann entropy is initially zero in all regimes, since the initial state is pure. It rises steeply from zero and attains higher values in the SSS and SWS regimes, where the bath is strongly coupled to the system. An interesting contrast emerges between strong and weak system-bath coupling scenarios. In the SSS and SWS regimes, the evolution of the entropy is qualitatively similar, whereas in the WSW and WWW regimes, higher values of entropy are attained when the inter-qubit coupling is weak. 

\subsubsection{Entropy production}
The irreversibility of a process is quantified using the entropy production $\Sigma$, which is a fundamental quantity that defines the second law of thermodynamics~\cite{Landi2021}. Consider the evolution of a composite system-bath $\rho_{SB} = U(t) \left[\rho_S(0)\otimes\rho_B(0)\right] U(t)^\dagger$, where $U(t)$ is the global unitary and $\rho_S(0)$ and $\rho_B(0)$ are the initial states of the system and the bath, respectively. Note that this evolution is general and is size-independent. Consequently, it can be applied, for example, to a system–bath configuration as small as two qubits, as well as to much larger setups. In general, the reduced state of the system of interest at any time $t$ is obtained by partially tracing the bath as $\rho_S(t) = {\rm Tr}_B\left\{U(t) \left[\rho_S(0)\otimes\rho_B(0)\right] U(t)^\dagger\right\}$. This partial tracing of the bath can be pointed out as the source of irreversibility. Thus, the process of discarding the information locally contained in the bath and discarding the non-local information shared between the system and the bath leads to irreversibility~\cite{Landi2021, Esposito_2010}. The definition of entropy production thus relies on these processes, and is given by
\begin{align}
    \Sigma(t) = \mathcal{I}\left[\rho_{SB}(t)\right] + S\left[\rho_B(t)\|\rho_B(0)\right],
\end{align}
where $\mathcal{I}[\rho_{SB}(t)] = S[\rho_S(t)] + S[\rho_B(t)] - S[\rho_{SB}(t)]$ [with $S[\rho]$ being the von Neumann entropy, Eq.~\eqref{von_Neumann_entropy}] is the mutual information developed between the system and the bath owing to the interaction between them, and $S[(*)||(\cdot)] = {\rm Tr}\left[(*)\log (*) - (*)\log (\cdot)\right]$ is the quantum relative entropy. After simplification, the above equation can be written as~\cite{Esposito_2010} 
\begin{align}
    \Sigma(t) = S[\rho_{SB}(t)\|\rho_S(t)\otimes\rho_B(0)],
\end{align}
which, if the bath is assumed to be initially in a thermal state $\rho_B^{\rm th}(0) = \exp(-\beta H_B)/{\rm Tr}\left[\exp(-\beta H_B)\right]$ at inverse temperature $\beta = 1/k_B T$, reduces to~\cite{Landi2021} 
\begin{align}\label{eq_sigma_thermal}
    \Sigma(t) = \Delta S[\rho_S(t)] + \beta Q_B(t),
\end{align}
where $\Delta S[\rho_S(t)] = S[\rho_S(t)] - S[\rho_S(0)]$ is the change in the von Neumann entropy and $Q_B(t) = {\rm Tr}\left[\{\rho_B(t) - \rho_B(0)\} H_B\right]$ is the change in the energy of the bath. In the above equation, the entropy production depends on the bath state, which is very difficult to determine in a situation such as ours.

However, in certain scenarios, such as in the presence of a global fixed point $U(t)\left[\rho^*_S\otimes \rho_B(0)\right]U(t)^\dagger = \rho_S^* \otimes \rho_B(0)$ or under the strict energy conservation condition $\left[U(t), H_S + H_B\right] = 0$, the entropy production can be obtained using only the system variables~\cite{Landi2021}. It can be readily verified that these conditions are not fulfilled for the two-qubit spin boson model considered here, Eq.~\eqref{eq_total_system_Hamiltonian}, in the arbitrary coupling regime, and therefore, we need the knowledge about the bath's variables to calculate the entropy production. Fortunately, the higher-level auxiliary density operators (ADOs) in the HEOM technique encode the system-bath correlations (see Appendix B) and can be used to obtain certain expectation values that contain bath variables~\cite{Tanimura_exact_2020, efficient_heom, tanimura_two_qubit_paper}. Thus, for example, using the HEOM, bath heat current for a system interacting with multiple bosonic baths has been calculated in~\cite{tanimura_heat_current}, and electronic current in the scenario of a system interacting with a fermionic bath has been calculated in~\cite{HierarchicalEOM}. 

In the two-qubit spin-boson model considered here, Eq.~\eqref{eq_total_system_Hamiltonian}, the system interacts with two baths at different temperatures. In such a scenario, entropy production, Eq.~\eqref{eq_sigma_thermal}, gets modified, and is given by 
\begin{align}\label{eq_sigma_thermal_2}
    \Sigma(t) = \Delta S\left[\rho_S(t)\right] + \beta_1Q_{B_1}(t) + \beta_2 Q_{B_2}(t). 
\end{align}
Calculating the change in bath energy directly using ADOs is numerically inefficient due to the presence of quadratic bath variables, see Eq.~\eqref{eq_total_system_Hamiltonian}. A possible approach is to represent the change in bath energy in terms of the system and interaction energies. Thus, the change in the bath energy, for non-interacting baths, can be written as $Q_{B_i}(t) = -Q_{S_i}(t) - Q_{SB_i}(t)$, where $Q_{S_i}(t) = i\int_0^t{\rm Tr}\left\{\left[H_{SB_i}, H_S\right]\rho_{SB}(s)\right\} ds$, and $Q_{SB_i}(t) = {\rm Tr}\left[\left\{\rho_{SB}(t) - \rho_{SB}(0)\right\}H_{SB_i}\right]$~\cite{tanimura_heat_current, tiwari_strong}. The interaction Hamiltonian is of the form $H_{SB_i} = \sigma_z^{(i)}\otimes B_i$, where $B_i$ is the linear sum of the bath harmonic oscillators. For a bosonic bath with Drude-Lorentz spectral density and correlation function $C_i(t) = \left\langle B_i(t)B_i(0)\right\rangle \simeq \sum_k c_{i, k}e^{-\nu_{i, k} t}$ (see Appendix B), the expectation value $Q_{SB_i} = \left\langle\sigma_z^{(i)}B_i(t)\right\rangle$, using HEOM, can be given by $\sum_k {\rm Tr}\left[\sigma_z^{(i)}\rho_1^{(i,k)}(t)\right]$, where $\rho_1^{(i, k)}$ is the first-level ADO~\cite{HierarchicalEOM}. Furthermore, the complexity in calculating the integration $Q_{S_i} = i\int_0^t{\rm Tr}\left\{\left[H_{SB_i}, H_S\right]\rho_{SB}(s)\right\} ds$ can be avoided by taking equal bath temperatures, such that $\beta_1 = \beta_2 = \beta$. In this case, using Eq.~\eqref{eq_total_system_Hamiltonian}, and after some algebraic manipulation, we can write 
\begin{align}\label{eq_change_in_bath_energy}
    \beta_1 Q_{B_1} + \beta_2 Q_{B_2} &= \beta\left(Q_{B_1} + Q_{B_2}\right)\nonumber \\
    &= -\beta\left(Q_S + Q_{SB_1} + Q_{SB_2}\right), 
\end{align}
where $Q_S = {\rm Tr}_S\left[\left\{\rho_S(t) - \rho_S(0)\right\}H_S\right]$ is the total change in the system's energy and depends entirely on the system variables. Using the above equation and Eq.~\eqref{eq_sigma_thermal_2}, a calculable form of entropy production for the two-qubit spin-boson model (with equal bath temperatures) is given by
\begin{align}\label{eq_final_ent_prod}
    \Sigma(t) = \Delta S[\rho_S(t)] - \beta\left[Q_S + Q_{SB_1} + Q_{SB_2}\right]. 
\end{align}

\begin{figure}
    \centering
    \includegraphics[width=1\linewidth]{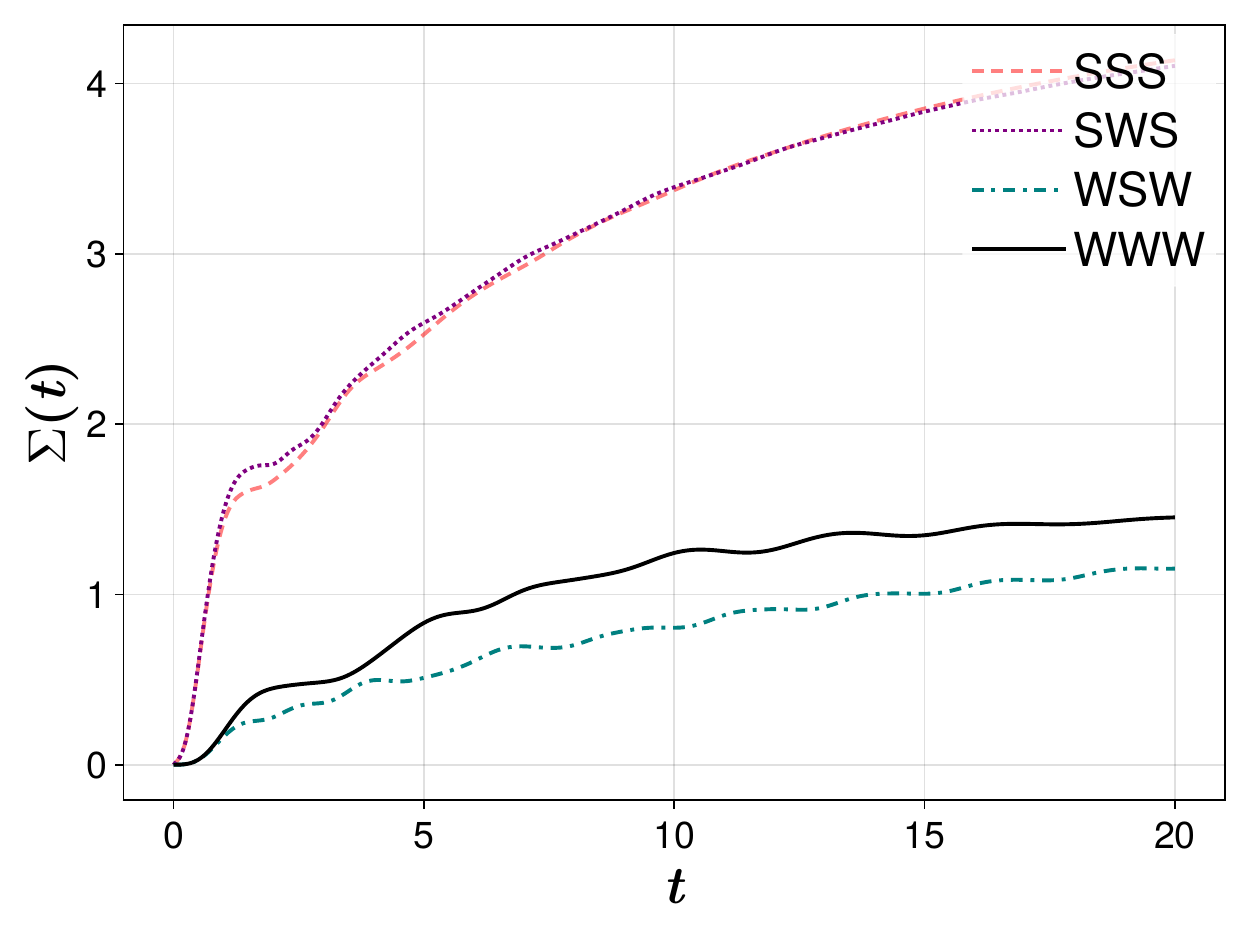}
    \caption{Variation of entropy production with time for the two-qubit spin-boson model in the SSS, SWS, WSW, and WWW coupling regimes. The initial state of the system is taken to be the excited state $\ket{00}_{12}$. The common bath temperature for qubits 1 and 2 is set to $T = 1.39$.}
    \label{fig_entropy_production}
\end{figure}

Figure~\ref{fig_entropy_production} depicts the variation of entropy production with time for the two-qubit spin-boson model, calculated using Eq.~\eqref{eq_final_ent_prod}, in the SSS, SWS, WSW, and WWW coupling regimes (see Sec.~\ref{setup}).
In all the regimes, the entropy production remains always positive, verifying the second law of quantum thermodynamics. The change in the von Neumann entropy of the system, shown in Fig.~\ref{fig_von_Neumann_entropy}, is one of the major components of the entropy production, Eq.~\eqref{eq_final_ent_prod}. The remaining component is the change in the baths' energy, Eq.~\eqref{eq_change_in_bath_energy}. Upon comparing Fig.~\ref{fig_entropy_production} with Fig.~\ref{fig_von_Neumann_entropy}, we observe an interesting yet consistent behavior. The initial-time behavior of the entropy production is similar to that of the von Neumann entropy. Furthermore, in the WWW and WSW regimes, the long-time behavior of entropy production exhibits a trend similar to that of the entropy of the system, in their relation to each other. A major deviation emerges in the SSS and SWS regimes, where the baths exert a stronger influence. In these regimes, changes in the baths' energy make a dominant contribution to the overall entropy production; therefore, the entropy production in these cases is higher than in the WWW and WSW regimes.

Having studied the dynamical changes in the thermodynamic properties of the two-qubit spin-boson model, we next explore the emergence of non-equilibrium steady state (NESS), its properties, and the relationship between heat and spin currents. 

\subsection{Non-equilibrium steady state (NESS) and the relationship between spin and heat currents}
A quantum system interacting with two baths at different temperatures at its edges is the simplest scenario of a non-equilibrium system. The steady state reached by the system in this case has a non-zero heat current flowing from one bath to another, and is therefore a non-equilibrium steady state (NESS)~\cite{Landi_ness_paper}. Here, in the case of the two-qubit spin-boson model, we have two distinct temperature baths interacting with the system at its edges, resulting in a NESS. To study the properties of the NESS corresponding to the two-qubit spin-boson model, we examine the steady-state coherence. The transport properties in the steady state, a hallmark of NESSs, are characterized by the spin and heat currents. Subsequently, the steady state of the two-qubit spin-boson model is obtained using the {\it steadystate} function of the {HierarchicalEOM.jl} package~\cite{HierarchicalEOM}. Additionally, in the next portion of the paper, we set the temperature of bath 1 ($T_1$), the cold bath in Fig.~\ref{fig_Schematic_two_qubit_spin_boson_model}, to one, and vary the temperature of bath 2 ($T_2 = T$), that is, the hot bath. 

\begin{figure}
    \centering
    \includegraphics[width=1\linewidth]{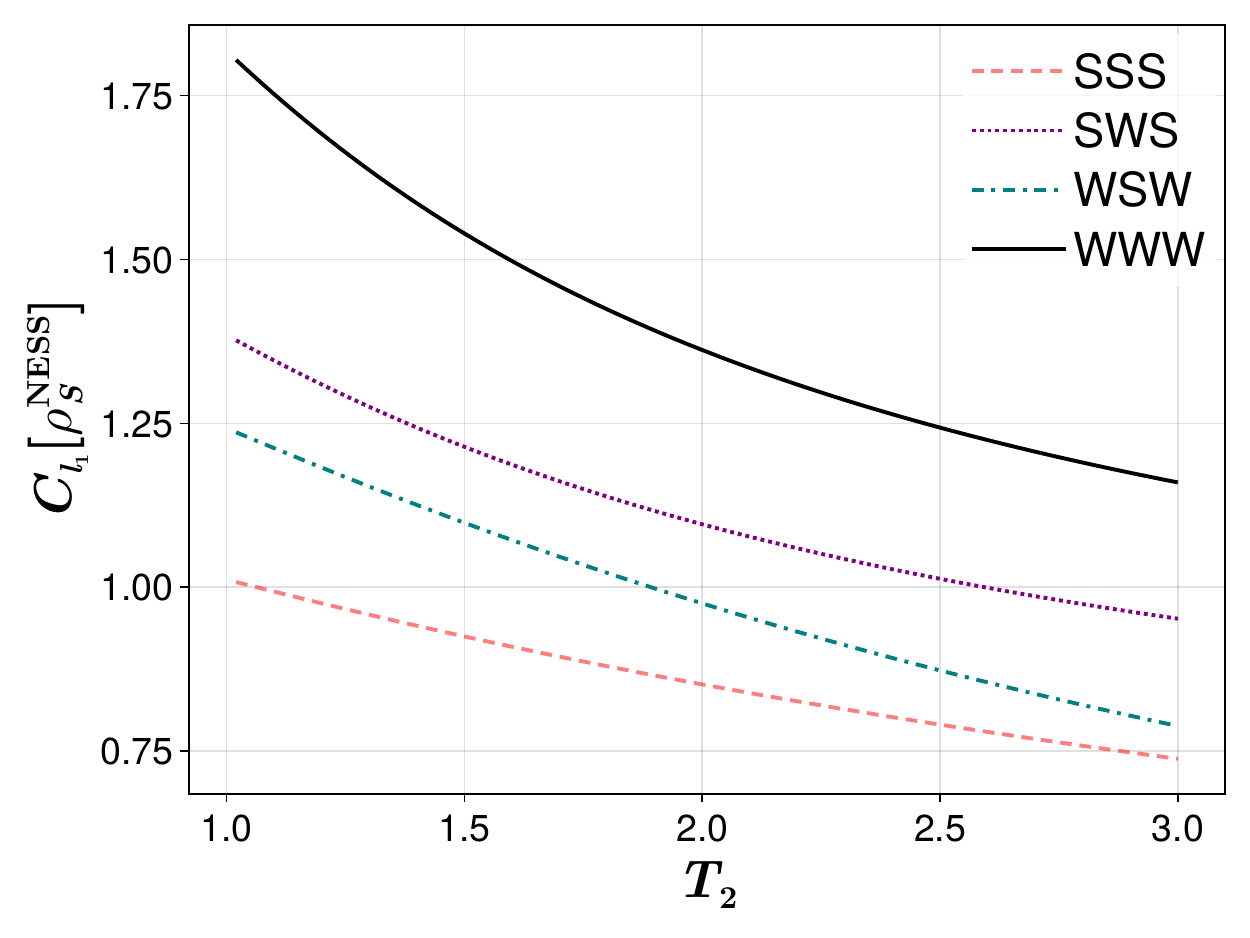}
    \caption{Variation of the $l_1$ norm of coherence, Eq.~\eqref{eq_l1_norm}, for the steady state of the two-qubit spin-boson model, $C_{l_1}\left[\rho_S^{\rm NESS}\right]$ with the temperature of the hot bath $T_2$ in various coupling regimes outlined in Sec.~\ref{setup}. Here, for numerical ease to get the steady state of the system, we have taken $\alpha_i = 1.5/\pi$ to denote strong system-bath coupling and $\alpha = 0.01/\pi$ for weak coupling. Similarly, for strong inter-qubit coupling strength $J = 1.5/\pi$ and $J = 0.01/\pi$ denotes weak inter-qubit coupling. The temperature of the cold bath, $T_1$, is fixed at 1.}
    \label{steady_state_coherence_with_T2}
\end{figure}
\subsubsection{Steady-state coherence and quantum correlations}
To examine the steady-state coherence of the two-qubit spin-boson model, we calculate the $l_1$ norm of coherence, defined in Eq.~\eqref{eq_l1_norm}, of the steady state of the system. This is illustrated in Fig.~\ref{steady_state_coherence_with_T2}.
It can be observed that the steady-state coherence decreases as the temperature of the hot bath increases. In the WWW regime, the steady-state coherence attains its highest values, while in the SSS coupling regime, it attains its lowest values. An interesting observation is made when the steady-state coherence in the WSW and the SWS regimes is compared. It is observed that when the system is strongly coupled to the baths, and the inter-qubit coupling is weak, the steady-state coherence is higher. Conversely, it is lower when the inter-qubit coupling is strong, and the system is weakly coupled to the baths.

\subsubsection{Quantum heat and spin currents}
The key characteristic of NESS is the presence of non-zero heat transport within the system in the steady state. To quantify this heat transport, we use the heat current. Consider the system Hamiltonian $H_S = H_1 + H_2 + H_{12}$ of the two-qubit spin-boson model, Eq.~\eqref{eq_total_system_Hamiltonian}. We compute the heat current between the two qubits of the system using the site-resolved continuity equation~\cite{Landi_ness_paper, Tiwari2025_circuit}. To this end, consider the rate of change in the energy of qubit 1
\begin{align}\label{eq_division_energy_heat}
    \frac{d}{dt}{\rm Tr}[H_1\rho_S(t)] = {\rm Tr}\left[\frac{dH_1}{dt}\rho_S(t)\right] + {\rm Tr}\left[H_1\frac{d\rho_S(t)}{dt}\right],
\end{align}
where the first term ${\rm Tr}\left[\frac{dH_1}{dt}\rho_S(t)\right]$ is the rate of work done and is zero as $H_1$ is time-independent. The second term is the rate of change in the heat of qubit 1. To proceed further, we need a master equation description of the system. This can be achieved using the principle of minimal dissipation~\cite{Hayden_2022, Colla_canHam_paper, tiwari2025_finite_bath}, where the master equation describing the dynamics of the system can be uniquely determined to be of the form $\frac{d\rho_S(t)}{dt} = \mathcal{L}[\rho_S(t)] = -i[H^{\rm can}(t), \rho_S(t)] + \mathcal{D}_{\rm min}[\rho_S(t)]$, where $H^{\rm can}(t)$ is the canonical Hamiltonian of the system and $\mathcal{D}_{\rm min}[\rho_S(t)]$ is the minimal dissipator. 

In general, the canonical Hamiltonian of the system can be decomposed as $H^{\rm can}(t) = H_S +$ {\it (time dependent bath's contribution to the Hermitian part)}, where the bath's contribution depends on how strongly the system is coupled to the bath and this contribution can be ignored when the system is weakly coupled to the bath~\cite{Colla_canHam_paper, Colla2025}. To get analytical closed form of the heat and spin currents and to observe the impact of tunneling amplitude and the inter-qubit coupling on them, we shall focus on the WWW and WSW regimes. On substituting the master equation in Eq.~\eqref{eq_division_energy_heat} for the rate of change in the heat of qubit 1, we get
\begin{align}\label{eq_division_system_and_bath_current}
    {\rm Tr}\left[H_1\frac{d\rho_S(t)}{dt}\right] = -i{\rm Tr}\left\{H_1\left[H_S, \rho_S(t)\right]\right\} + {\rm Tr}\left\{H_1\mathcal{D}[\rho_S(t)]\right\},
\end{align}
where the first term on the right side of the equation is the contribution to the rate of change in the heat of qubit 1, that is, {\it heat current}, from inter-qubit coupling with qubit 2, and the second term is the contribution from the corresponding bath. 

Here, we analyze the heat current flowing from qubit 2 to qubit 1 $\mathcal{J}_{21}$ in the NESS regime for the case where the system is weakly coupled to the bath. In the steady-state regime, the left-hand side of Eq.~\eqref{eq_division_system_and_bath_current} vanishes. Consequently, the total heat current entering and leaving qubit 1 (and analogously for qubit 2) becomes zero, which represents the quantum thermal analogue of {\it Kirchhoff's current law}~\cite{Tiwari2025_circuit}. This implies that the heat current entering qubit 1 from qubit 2 is equal and opposite to the heat current leaving it. Using Eqs.~\eqref{eq_total_system_Hamiltonian} and~\eqref{eq_division_system_and_bath_current}, and after simplification, the heat current $\mathcal{J}_{21}$ is given by 
\begin{align}\label{eq_heat_current_final}
    \mathcal{J}_{21} &= i\left\langle\left[H_{12}, H_1\right]\right\rangle \nonumber \\
    &= \Delta_1 J\left\langle\sigma_z^{(1)}\sigma_y^{(2)}\right\rangle - \epsilon_1J\left\langle\sigma^{(1)}_x\sigma_y^{(2)} - \sigma_y^{(1)}\sigma_x^{(2)}\right\rangle,
\end{align}
where $\braket{(\cdot)} = {\rm Tr}\left[(\cdot)\rho_S^{\rm NESS}\right]$. This heat current is a central quantity in the design of quantum thermal analogs of electric circuits, such as quantum thermal diodes, transistors, adders, and Wheatstone bridges~\cite{Tiwari2025_circuit}.

In a similar setup, the spin current is also talked about~\cite{Poletti_2018, Poulsen_2022, Poulsen_2024, Stephen_clark_2024}, which can be experimentally measured~\cite{Landi_ness_paper, Maekawa_2013, Bloch_2016}. The spin current, between qubits 1 and 2, can be obtained using the spin-continuity equation~\cite{Landi2014} 
\begin{align}
    \frac{d}{dt}\left\langle\sigma_z^{(1)}\right\rangle = j_{c1} - j_{12},
\end{align}
where $j_{c1}$ is the flux from the cold bath to qubit 1, and $j_{12}$ is the flux outgoing from qubit 1 to qubit 2. On performing a similar derivation as for the heat current, we end up with the following form of the spin current $j_{12}$
\begin{align}\label{eq_spin_current_final}
    j_{12} &= -i \left\langle\left[H_{12}, \sigma_z^{(1)}\right]\right\rangle \nonumber \\
    &= -\Delta_1\left\langle\sigma_y^{(1)}\right\rangle + 2J\left\langle\sigma_x^{(1)}\sigma_y^{(2)} - \sigma_y^{(1)}\sigma_x^{(2)}\right\rangle.
\end{align}
On comparing the above equation with Eq.~\eqref{eq_heat_current_final}, it can be readily observed that setting the tunneling amplitude to $\Delta_1 = 0$ yields a direct relationship between the heat and spin currents (in the weak coupling regime), that is,
\begin{align}
    |\mathcal{J}_{21}| = \frac{\epsilon_1}{2}|j_{12}|.
\end{align}
This highlights that determining either the spin current or the heat current immediately provides information about the other. However, in the spin-boson model, the tunneling amplitude $\Delta_1$ is an essential parameter of the system, and it disrupts the direct correspondence between the heat and spin currents. Note that a similar set of equations for heat and spin currents can be obtained by keeping qubit 2 as a node.   

\section{Conclusions}\label{sec_conclusions}
In this work, we studied a two-qubit spin-boson model in the strong-coupling regime. Two non-perturbative techniques were employed to obtain the dynamics of the two-qubit system, namely the reaction-coordinate mapping and the hierarchical equation of motion. Both techniques were compared and found to be in tandem. The spin-boson model was characterized by four different coupling regimes, based on the strength of interaction with the baths and the inter-qubit interaction strength. To characterize the quantum nature of the evolution, quantum coherence and non-Markovianity were investigated. Quantum coherence (studied using the $l_1$ norm of coherence) emerged over the course of the evolution, even though the initial state contained no coherence. The tunneling amplitude was crucial in determining the degree of generated coherence. In general, larger tunneling amplitudes produced the higher peaks in the $l_1$ norm, whereas smaller tunneling amplitudes resulted in consistently lower values. A weak system-bath coupling and strong inter-qubit coupling were found to be the favorable scenario for maximum coherence generation and its sustainment. Quantum non-Markovianity was investigated using the trace distance measure. The oscillatory evolution of the trace distance demonstrated the non-Markovian nature of the system's evolution. The non-Markovianity strength was found to be greater in the scenario with weak inter-qubit coupling and strong system-bath coupling. 

We next explored the thermodynamic features of the model. To this end, we studied the von Neumann entropy and entropy production, thereby examining the second law of thermodynamics. In general, the von Neumann entropy took higher values when the system was strongly coupled to the baths. The entropy production was consistently higher for strong system-bath couplings and was always positive across all regimes, consistent with the second law of thermodynamics. Furthermore, the non-equilibrium steady-state properties of the system were probed using steady-state coherence, heat, and spin current. The steady-state coherence was maximum for weak system-bath and inter-qubit couplings. Additionally, the steady-state coherence monotonically decreased as the temperature difference of the two baths increased. Next, an important relationship between the spin and heat currents was demonstrated in the scenario of weak system-bath coupling. In general, the literature shows that, for Markovian evolution and zero tunneling amplitude, the spin current is linearly related to the heat current. This has also been demonstrated here. However, this work demonstrated that both the spin and heat currents depend on the tunneling amplitude, and that for non-zero tunneling, the existing relationship doesn't hold. Collectively, these insights into the interplay between coherence, non-Markovianity, and thermodynamic fluxes provide a robust blueprint for engineering the next generation of quantum thermal machines operating beyond the weak-coupling limit.

\appendix

\section{The Reaction-Coordinate Mapping Approach}
Following Refs.~\cite{IlesSmith2014} and~\cite{Strasberg2016}, we perform the following RC mapping:\\

\noindent
\textit{For the cold bath interacting with qubit 1:}
\begin{align}
        \sum_{k}g_{k}q_{k} = \lambda_{1}Q_{1}, &&
        \vec{Q} = \Lambda^{(1)}\vec{q}, &&
        \vec{P} = \Lambda^{(1)}\vec{p},
\end{align}
where the orthogonal transformation matrix \(\Lambda^{(1)}\) follows,
\begin{equation}
    \sum_{j = 1}^{n}\Lambda^{(1)}_{kj}\Lambda^{(1)}_{lj} = \delta_{kl}.
\end{equation}
Thus, the transformation equations are
\begin{equation}
    \begin{aligned}
        Q_{1} &= \sum_{k}\Lambda_{1k}^{(1)}q_{k}, \quad & P_{1} &= \sum_{k}\Lambda_{1k}^{(1)}p_{k}, \\
        Q_{j} &= \sum_{k}\Lambda_{jk}^{(1)}q_{k}, \quad & P_{j} &= \sum_{k}\Lambda_{jk}^{(1)}p_{k},
    \end{aligned}
\end{equation}
and inverse transformations are
\begin{align}
    q_{k} &= \sum_{j}\Lambda_{jk}^{(1)}Q_{k}, && p_{k} = \sum_{j}\Lambda_{jk}^{(1)}P_{j}.
\end{align}

\noindent
\textit{For the hot bath interacting with qubit 2:} 
\begin{align}
        \sum_{l}h_{l}x_{l} &= \lambda_{2}X_{1}, &&
        \vec{X} = \Lambda^{(2)}\vec{x}, &&
        \vec{\Pi} = \Lambda^{(2)}\vec{\pi},
\end{align}
where the orthogonal transformation matrix \(\Lambda^{(2)}\) follows
\begin{equation}
    \sum_{j = 1}^{n}\Lambda^{(2)}_{kj}\Lambda^{(2)}_{lj} = \delta_{kl}. 
\end{equation}
In this case, the transformation equations become
\begin{align}
        X_{1} &= \sum_{l}\Lambda_{1l}^{(2)}x_{l}, && \Pi_{1} = \sum_{l}\Lambda_{1l}^{(2)}\pi_{l}, \\
        X_{m} &= \sum_{l}\Lambda_{ml}^{(2)}x_{l}, && \Pi_{m} = \sum_{l}\Lambda_{ml}^{(2)}\pi_{l},
\end{align}
and the corresponding inverse transformations become
\begin{align}
    x_{l} &= \sum_{m}\Lambda_{ml}^{(2)}X_{m}, &&
    \pi_{l} = \sum_{m}\Lambda_{ml}^{(2)}\Pi_{m}. 
\end{align}

Applying the above-mentioned transformations in the total Hamiltonian, Eq.~\eqref{eq_total_system_Hamiltonian}, we find the following RC-mapped Hamiltonian
\begin{align}
    H_{RC} &= \sum_{i = 1}^{2} \left( \frac{\epsilon_{i}}{2} \sigma_{z}^{(i)} + \frac{\Delta_{i}}{2} \sigma_{x}^{(i)} \right) 
    + J \left( \sigma_{x}^{(1)} \sigma_{x}^{(2)} + \sigma_{y}^{(1)} \sigma_{y}^{(2)} \right) \nonumber \\
    &\quad + \frac{1}{2} \left( \Pi_{1}^{2} + \Gamma_{1}^{2} X_{1}^{2} \right) 
    + \frac{1}{2} \left( P_{1}^{2} + \Omega_{1}^{2} Q_{1}^{2} \right) 
    + \lambda_{1} \sigma_{z}^{(1)} Q_{1} \nonumber \\
    &+ \lambda_{2} \sigma_{z}^{(2)} X_{1} + \sum_{k \neq 1} \frac{1}{2} \left( P_{k}^{2} + \Omega_{k}^{2} Q_{k}^{2} \right) 
    + \sum_{l \neq 1} \frac{1}{2} \left( \Pi_{l}^{2} + \Gamma_{l}^{2} X_{l}^{2} \right) \nonumber \\
    &\quad - Q_{1} \sum_{k \neq 1} C_{k} Q_{k} - X_{1} \sum_{l \neq 1} D_{l} X_{l},
\end{align}
where \( Q_1, P_1 \) are the position and momentum operators of the reaction coordinate (RC) mode associated with the bath coupled to qubit 1, \( X_1, \Pi_1 \) are the position and momentum operators of the RC mode associated with the bath coupled to qubit 2, \( \Omega_1, \Gamma_1 \) are the effective frequencies of the RC modes, \( \lambda_1 \) is the coupling strength between qubit 1 and its corresponding RC mode \(Q_1\), and \( \lambda_2 \) is the coupling strength between qubit 2 and its associated RC mode \(X_1\). Furthermore, \( Q_k, P_k \) are the position and momentum operators of the residual bath modes (not part of RC) originally associated with qubit 1, \( X_l, \Pi_l \) are the position and momentum operators of the residual bath modes originally associated with qubit 2, and \( \Omega_k, \Gamma_l \) are the frequencies of the residual modes from baths 1 and 2, respectively. Also, \( C_k \) is the coupling strength between the RC mode \( Q_1 \) and its residual bath modes \( Q_k \) and \( D_l \) is the coupling strength between the RC mode \( X_1 \) and its residual bath modes \( X_l \).

In terms of annihilation and creation operators, $c_1, d_1$ and $c_1^\dagger, d_1^\dagger$, respectively, the operators $Q_1, P_1$ and $X_1, \Pi_1$ are written as
\begin{equation}
\begin{aligned}
    Q_{1} &= \sqrt{\frac{\hbar}{2\Omega_{1}}} \left( c_{1}^{\dagger} + c_{1} \right), 
    &\quad P_{1} &= i \sqrt{\frac{\hbar\Omega_{1}}{2}} \left( c_{1}^{\dagger} - c_{1} \right), \\
    X_{1} &= \sqrt{\frac{\hbar}{2\Gamma_{1}}} \left( d_{1}^{\dagger} + d_{1} \right), 
    &\quad \Pi_{1} &= i \sqrt{\frac{\hbar\Gamma_{1}}{2}} \left( d_{1}^{\dagger} - d_{1} \right), \label{11}
\end{aligned}
\end{equation}
and the form of supersystem Hamiltonian \(H_{\rm sys}^{\prime}\) is given by
\begin{align}\label{eq_appA_Hsys_prime}
    H_{\text{sys}}' &= \sum_{i = 1}^{2} \left( \frac{\epsilon_{i}}{2} \sigma_{z}^{(i)} + \frac{\Delta_{i}}{2} \sigma_{x}^{(i)} \right) 
    + J \left( \sigma_{x}^{(1)} \sigma_{x}^{(2)} + \sigma_{y}^{(1)} \sigma_{y}^{(2)} \right) \nonumber \\
    &+ \frac{1}{2} \left( \Pi_{1}^{2} + \Gamma_{1}^{2} X_{1}^{2} \right) + \frac{1}{2} \left( P_{1}^{2} + \Omega_{1}^{2} Q_{1}^{2} \right) + \lambda_{1} \sigma_{z}^{(1)} Q_{1} \nonumber \\
    &+ \lambda_{2} \sigma_{z}^{(2)} X_{1}.
\end{align}
This form will be important in numerical implementation.

For the case of two qubits interacting with their respective bosonic baths, the interaction Hamiltonian has the form
\begin{align}
    H_{I} &= S_{1}\otimes B_{1} + S_{2}\otimes B_{2} \nonumber \\
    &= -Q_{1}\otimes\sum_{k \neq 1}C_{k}Q_{k} - X_{1}\otimes\sum_{l \neq 1}D_{l}X_{l}.
\end{align}
Also, in the interaction picture, the RC operators are given by
\begin{align}
        S_{1}(-\tau) &= \sum_{kl}e^{-i\Omega_{kl}\tau}S_{1}^{kl} |k\rangle\langle l|, \nonumber \\
        S_{2}(-\tau) &= \sum_{kl}e^{-i\Gamma_{kl}\tau}S_{2}^{kl} |k\rangle\langle l|, 
\end{align}
where \(|k\rangle\) represents the energy eigenstate of the super-systen Hamiltonian, \(H_{\rm sys}^{\prime}\). Following the procedure outlined in~\cite{IlesSmith2014}, we find the RC-mapped master equation of the two-qubit spin-boson model to be 
\begin{equation}
    \frac{d}{dt}\rho(t) = -i\big[H_{\rm sys}^{\prime}, \rho(t)\big] 
    - \sum_{\nu = 1}^{2}\Big[S_{\nu}, \big[\chi_{\nu}, \rho(t)\big]\Big] 
    + \sum_{\nu = 1}^{2}\Big[S_{\nu}, \big\{\Theta_{\nu}, \rho(t)\big\}\Big],
\end{equation}
where
\begin{align}
        \chi_{1} &= \frac{1}{2}\sum_{\Omega_{kl}}J^{(1)}_{RC}~\Omega_{kl} \coth\left(\frac{\beta_{1}\Omega_{kl}}{2}\right)S_{1}^{kl}|k\rangle\langle l| \nonumber \\
        \Theta_{1} &= \frac{1}{2}\sum_{\Omega_{kl}}J^{(1)}_{RC}~\Omega_{kl}S_{1}^{kl} |k\rangle\langle l| \\
        \chi_{2} &= \frac{1}{2}\sum_{\Gamma_{kl}}J^{(2)}_{RC}~\Gamma_{kl}\coth\left(\frac{\beta_{2}\Gamma_{kl}}{2}\right)S_{2}^{kl}|k\rangle\langle l| \\
        \Theta_{2} &= \frac{1}{2}\sum_{\Gamma_{kl}}J^{(2)}_{RC}~\Gamma_{kl}S_{2}^{kl} |k\rangle\langle l|.
\end{align}
Here, \(H_{sys}^{\prime}\) takes the form given in Eq.~\eqref{eq_appA_Hsys_prime} and \( J^{(1)}_{RC}(\omega) \) and \( J^{(2)}_{RC}(\omega) \) denote the post RC mapping spectral densities corresponding to the coupling of qubits 1 and 2 with their respective baths. These take an Ohmic form if the pre–RC mapping spectral densities are assumed to be of Drude–Lorentz type~\cite{IlesSmith2014}. Here, we take the pre-mapping spectral densities to be of Drude-Lorentz type for each bath, given by
\begin{align}
    J^{(1)}(\omega) &= \frac{\alpha_{1}\omega\omega_{c_1}}{\omega^2 + \omega_{c_1}^2}, &&
    J^{(2)}(\omega) = \frac{\alpha_{2}\omega\omega_{c_2}}{\omega^2 + \omega_{c_2}^2}, 
    \label{eq_appA_spectral_density}
\end{align}
where $\omega_{c_1}$ and $\omega_{c_2}$ are the corresponding cut-off frequencies.

\section{The Hierarchical Equation of Motion Approach}

Here, we revisit the total Hamiltonian of the two-qubit spin boson model given by
\begin{equation}
H_{\mathrm{tot}}
= H_S + \sum_{i=1}^2 \left( H_{B,i} + H_{SB,i} \right),
\end{equation}
where $i=1,2$ labels the two baths, and the system Hamiltonian is given by~\eqref{eq_total_system_Hamiltonian} 
\begin{equation}
\begin{aligned}
H_S &= \frac{\epsilon_1}{2}\sigma_1^z + \frac{\Delta_1}{2}\sigma_1^x
     + \frac{\epsilon_2}{2}\sigma_2^z + \frac{\Delta_2}{2}\sigma_2^x  \\
&\quad + J\left(\sigma_1^x \sigma_2^x + \sigma_1^y \sigma_2^y \right).
\end{aligned}
\end{equation}
Each qubit couples longitudinally to its own bosonic environment via
\begin{equation}
H_{SB,i} = \sigma_i^z \otimes B_i,
\qquad 
B_i = \sum_k g_{k,i} (a_{k,i} + a_{k,i}^\dagger),
\end{equation}
with the bath Hamiltonian
\begin{equation}
H_{B,i} = \sum_k \omega_{k,i} a_{k,i}^\dagger a_{k,i}.
\end{equation}
Here, we consider the Drude-Lorentz spectral density
\begin{equation}
J_i(\omega) = \frac{\lambda_i \, \omega \, \Omega_i}{\omega^2 + \Omega_i^2},
\end{equation}
where $\lambda_i$ denotes the coupling between the system and the respective bath and $\Omega_i$ is the cutoff frequency. On comparing with Eq.~\eqref{eq_appA_spectral_density}, we have $\alpha_i = \lambda_i$ and $\Omega_i = \omega_{c_i}$.   
At temperature $T_i$, the bath correlation function is given by
\begin{equation}
C_i(t) = \langle B_i(t)B_i(0)\rangle_{T_i},
\end{equation}
which admits decomposition of the form
\begin{equation}
C_i(t) \approx \sum_{k=0}^{K_i} c_{i,k} \, e^{-\nu_{i,k} t},
\end{equation}
where the coefficients $c_{i,k}$ and decay rates $\nu_{i,k}$ arise from
the Drude pole and from a Pad\'e or Matsubara expansion of the Bose-Einstein function.

The Hierarchical equation of motion (HEOM) method introduces a hierarchy of auxiliary density operators (ADOs)
$\{\rho_{\vec{n}}(t)\}$ indexed by a multi--index $\vec{n}=\{n_{i,k}\}$ associated with the exponential expansion of the bath correlations.
The physical reduced density matrix is the zeroth-tier ADO
\begin{equation}
\rho_{\vec{0}}(t) \equiv \rho_S(t) = \mathrm{Tr}_B\big[\rho_{\mathrm{tot}}(t)\big].
\end{equation}
The equations of motion for the ADOs take the form
\begin{align}
\dot{\rho}_{\vec{n}}
 &= -i[H_S,\rho_{\vec{n}}]
 - \sum_{i,k} n_{i,k} \nu_{i,k} \rho_{\vec{n}}  - i \sum_{i,k} 
 \left( Q_i \rho_{\vec{n}+\hat{e}_{i,k}}
      - \rho_{\vec{n}+\hat{e}_{i,k}} Q_i \right) \nonumber \\
 & - i \sum_{i,k} n_{i,k} c_{i,k}
 \left( Q_i \rho_{\vec{n}-\hat{e}_{i,k}}
      - \rho_{\vec{n}-\hat{e}_{i,k}} Q_i \right),
\label{eq:heom}
\end{align}
where $\hat{e}_{i,k}$ increments the $(i,k)$th component of the hierarchy index and $Q_i$ is the system operator. The hierarchy is truncated by imposing a maximum depth
\begin{equation}
\sum_{i,k} n_{i,k} \le L,
\end{equation}
with $L$ chosen large enough to ensure convergence of system observables. Each bath is assumed initially thermal
\begin{equation}
\rho_{B,i}(0) = \frac{e^{-H_{B,i}/k_B T_i}}{Z_i}.
\end{equation}
Under the usual factorized initial condition,
\begin{equation}
\rho_{\mathrm{tot}}(0) = \rho_S(0) \otimes \rho_{B,1}(0) \otimes \rho_{B,2}(0),
\end{equation}
the HEOM hierarchy is initialized as
\begin{equation}
\rho_{\vec{0}}(0) = \rho_S(0),
\qquad
\rho_{\vec{n}}(0) = 0
\quad (\vec{n}\ne \vec{0}),
\end{equation}
since the effects of the thermal bath states are fully encoded through their correlation functions in the dynamical coefficients of the hierarchy.

After time propagation of Eq.~(\ref{eq:heom}), the physical reduced density matrix at time $t$ is given directly by the zeroth-tier ADO $\rho_{\vec{0}}(t)$.
Expectation values of any system operator $O$ follow from
\begin{equation}
\langle O(t) \rangle 
= \mathrm{Tr}\!\left[ \rho_{\vec{0}}(t)\, O \right].
\end{equation}
Reduced states of individual qubits are found by partial tracing
\begin{equation}
\rho_{(1)}(t) = \mathrm{Tr}_2\!\left[\rho_{\vec{0}}(t)\right],
\qquad
\rho_{(2)}(t) = \mathrm{Tr}_1\!\left[\rho_{\vec{0}}(t)\right].
\end{equation}
Higher-tier ADOs encode non-Markovian memory and system-bath correlations but are not themselves physical density operators. The HEOM framework provides a numerically exact treatment of system-bath interactions (up to hierarchy truncation) and is capable of accurately describing strong coupling and finite temperatures, making it well-suited to investigating coherent and dissipative dynamics in coupled qubit systems. In the numerical simulations using {\it HierarchicalEOM.jl}~\cite{HierarchicalEOM} and {\it QuantumToolbox.jl}~\cite{QuantumToolbox}, we retain around four terms in the Pad\'e expansion of each bath correlation function and truncate the hierarchy at tier $L=6$.

\bibliographystyle{apsrev4-2}
\bibliography{reference}

@book{Breuer2002,
    author = {Breuer, Heinz-Peter and Petruccione, Francesco},
    title = {The Theory of Open Quantum Systems},
    publisher = {Oxford University Press},
    year = {2007},
    month = {01},
    isbn = {9780199213900},
    doi = {10.1093/acprof:oso/9780199213900.001.0001},
    url = {https://doi.org/10.1093/acprof:oso/9780199213900.001.0001},
}

@article{IlesSmith2014,
  title = {Environmental dynamics, correlations, and the emergence of noncanonical equilibrium states in open quantum systems},
  author = {Iles-Smith, Jake and Lambert, Neill and Nazir, Ahsan},
  journal = {Phys. Rev. A},
  volume = {90},
  issue = {3},
  pages = {032114},
  numpages = {9},
  year = {2014},
  month = {Sep},
  publisher = {American Physical Society},
  doi = {10.1103/PhysRevA.90.032114},
  url = {https://link.aps.org/doi/10.1103/PhysRevA.90.032114}
}

@article{Strasberg2016,
doi = {10.1088/1367-2630/18/7/073007},
url = {https://doi.org/10.1088/1367-2630/18/7/073007},
year = {2016},
month = {jul},
publisher = {IOP Publishing},
volume = {18},
number = {7},
pages = {073007},
author = {Strasberg, Philipp and Schaller, Gernot and Lambert, Neill and Brandes, Tobias},
title = {Nonequilibrium thermodynamics in the strong coupling and non-Markovian regime based on a reaction coordinate mapping},
journal = {New Journal of Physics}
}

@article{Tiwari2025_circuit,
  title = {Quantum Thermal Analogs of Electric Circuits: A Universal Approach},
  author = {Tiwari, Devvrat and Bhattacharya, Samyadeb and Banerjee, Subhashish},
  journal = {Phys. Rev. Lett.},
  volume = {135},
  issue = {2},
  pages = {020404},
  numpages = {6},
  year = {2025},
  month = {Jul},
  publisher = {American Physical Society},
  doi = {10.1103/5x8m-bhgd},
  url = {https://link.aps.org/doi/10.1103/5x8m-bhgd}
}

@article{CaldeiraLeggett1981,
title = {Quantum tunnelling in a dissipative system},
journal = {Annals of Physics},
volume = {149},
number = {2},
pages = {374-456},
year = {1983},
issn = {0003-4916},
doi = {https://doi.org/10.1016/0003-4916(83)90202-6},
url = {https://www.sciencedirect.com/science/article/pii/0003491683902026},
author = {A.O Caldeira and A.J Leggett}
}

@article{Leggett1987,
  title = {Dynamics of the dissipative two-state system},
  author = {Leggett, A. J. and Chakravarty, S. and Dorsey, A. T. and Fisher, Matthew P. A. and Garg, Anupam and Zwerger, W.},
  journal = {Rev. Mod. Phys.},
  volume = {59},
  issue = {1},
  pages = {1--85},
  numpages = {0},
  year = {1987},
  month = {Jan},
  publisher = {American Physical Society},
  doi = {10.1103/RevModPhys.59.1},
  url = {https://link.aps.org/doi/10.1103/RevModPhys.59.1}
}

@article{Landi2021,
  title = {Irreversible entropy production: From classical to quantum},
  author = {Landi, Gabriel T. and Paternostro, Mauro},
  journal = {Rev. Mod. Phys.},
  volume = {93},
  issue = {3},
  pages = {035008},
  numpages = {58},
  year = {2021},
  month = {Sep},
  publisher = {American Physical Society},
  doi = {10.1103/RevModPhys.93.035008},
  url = {https://link.aps.org/doi/10.1103/RevModPhys.93.035008}
}

@article{BLP_2009,
  title = {Measure for the Degree of Non-Markovian Behavior of Quantum Processes in Open Systems},
  author = {Breuer, Heinz-Peter and Laine, Elsi-Mari and Piilo, Jyrki},
  journal = {Phys. Rev. Lett.},
  volume = {103},
  issue = {21},
  pages = {210401},
  numpages = {4},
  year = {2009},
  month = {Nov},
  publisher = {American Physical Society},
  doi = {10.1103/PhysRevLett.103.210401},
  url = {https://link.aps.org/doi/10.1103/PhysRevLett.103.210401}
}

@article{Landi2014,
  title = {Flux rectification in the quantum $XXZ$ chain},
  author = {Landi, Gabriel T. and Novais, E. and de Oliveira, M\'ario J. and Karevski, Dragi},
  journal = {Phys. Rev. E},
  volume = {90},
  issue = {4},
  pages = {042142},
  numpages = {11},
  year = {2014},
  month = {Oct},
  publisher = {American Physical Society},
  doi = {10.1103/PhysRevE.90.042142},
  url = {https://link.aps.org/doi/10.1103/PhysRevE.90.042142}
}

@article{Chen2020_polaron,
  title = {Full-polaron master equation approach to dynamical steady states of a driven two-level system beyond the weak system-environment coupling},
  author = {Chen, Chien-Chang and Stace, Thomas M. and Goan, Hsi-Sheng},
  journal = {Phys. Rev. B},
  volume = {102},
  issue = {3},
  pages = {035306},
  numpages = {9},
  year = {2020},
  month = {Jul},
  publisher = {American Physical Society},
  doi = {10.1103/PhysRevB.102.035306},
  url = {https://link.aps.org/doi/10.1103/PhysRevB.102.035306}
}

@article{Esposito2015_NEGF,
  title = {Quantum Thermodynamics: A Nonequilibrium Green's Function Approach},
  author = {Esposito, Massimiliano and Ochoa, Maicol A. and Galperin, Michael},
  journal = {Phys. Rev. Lett.},
  volume = {114},
  issue = {8},
  pages = {080602},
  numpages = {5},
  year = {2015},
  month = {Feb},
  publisher = {American Physical Society},
  doi = {10.1103/PhysRevLett.114.080602},
  url = {https://link.aps.org/doi/10.1103/PhysRevLett.114.080602}
}

@article{HierarchicalEOM,
  title = {An efficient {J}ulia framework for hierarchical equations of motion in open quantum systems},
  author = {Huang, Yi-Te and Kuo, Po-Chen and Lambert, Neill and Cirio, Mauro and Cross, Simon and Yang, Shen-Liang and Nori, Franco and Chen, Yueh-Nan},  
  journal = {Communications Physics},
  publisher = {Nature Portfolio},
  volume = {6},
  number = {1},
  pages = {313},
  month = {Oct},
  year = {2023},
  doi = {10.1038/s42005-023-01427-2},
  url = {https://doi.org/10.1038/s42005-023-01427-2}
}

@article{QuantumToolbox,
  title = {Quantum{T}oolbox.jl: {A}n efficient {J}ulia framework for simulating open quantum systems},
  author = {Mercurio, Alberto and Huang, Yi-Te and Cai, Li-Xun and Chen, Yueh-Nan and Savona, Vincenzo and Nori, Franco},
  journal = {{Quantum}},
  issn = {2521-327X},
  publisher = {{Verein zur F{\"{o}}rderung des Open Access Publizierens in den Quantenwissenschaften}},
  volume = {9},
  pages = {1866},
  month = sep,
  year = {2025},
  doi = {10.22331/q-2025-09-29-1866},
  url = {https://doi.org/10.22331/q-2025-09-29-1866}
}

@article{Streltsov_2017,
  title = {Colloquium: Quantum coherence as a resource},
  author = {Streltsov, Alexander and Adesso, Gerardo and Plenio, Martin B.},
  journal = {Rev. Mod. Phys.},
  volume = {89},
  issue = {4},
  pages = {041003},
  numpages = {34},
  year = {2017},
  month = {Oct},
  publisher = {American Physical Society},
  doi = {10.1103/RevModPhys.89.041003},
  url = {https://link.aps.org/doi/10.1103/RevModPhys.89.041003}
}

@article{Baumgratz_2014,
  title = {Quantifying Coherence},
  author = {Baumgratz, T. and Cramer, M. and Plenio, M. B.},
  journal = {Phys. Rev. Lett.},
  volume = {113},
  issue = {14},
  pages = {140401},
  numpages = {5},
  year = {2014},
  month = {Sep},
  publisher = {American Physical Society},
  doi = {10.1103/PhysRevLett.113.140401},
  url = {https://link.aps.org/doi/10.1103/PhysRevLett.113.140401}
}

@misc{kading2025,
      title={Density matrices in quantum field theory: Non-Markovianity, path integrals and master equations}, 
      author={Christian Käding and Mario Pitschmann},
      year={2025},
      eprint={2503.08567},
      archivePrefix={arXiv},
      primaryClass={hep-th},
      url={https://arxiv.org/abs/2503.08567}, 
}

@article{Rivas_2014, 
title={Quantum non-Markovianity: characterization, quantification and detection}, 
volume={77}, 
ISSN={1361-6633}, 
url={http://dx.doi.org/10.1088/0034-4885/77/9/094001}, 
DOI={10.1088/0034-4885/77/9/094001}, 
number={9}, 
journal={Reports on Progress in Physics}, 
publisher={IOP Publishing}, 
author={Rivas, \'Angel and Huelga, Susana F and Plenio, Martin B}, 
year={2014}, 
month=aug, 
pages={094001} 
}

@article{Utagi_2020, 
author={Utagi, Shrikant
and Srikanth, R.
and Banerjee, Subhashish},
title={Temporal self-similarity of quantum dynamical maps as a concept of memorylessness},
journal={Scientific Reports},
year={2020},
month={Sep},
day={14},
volume={10},
number={1},
pages={15049},
issn={2045-2322},
doi={10.1038/s41598-020-72211-3},
url={https://doi.org/10.1038/s41598-020-72211-3}
}

@article{de_Vega_2017, 
  title = {Dynamics of non-Markovian open quantum systems},
  author = {de Vega, In\'es and Alonso, Daniel},
  journal = {Rev. Mod. Phys.},
  volume = {89},
  issue = {1},
  pages = {015001},
  numpages = {58},
  year = {2017},
  month = {Jan},
  publisher = {American Physical Society},
  doi = {10.1103/RevModPhys.89.015001},
  url = {https://link.aps.org/doi/10.1103/RevModPhys.89.015001}
}

@BOOK{Nielsen2010,
  title     = "Quantum computation and quantum information",
  author    = "Nielsen, Michael A and Chuang, Isaac L",
  publisher = "Cambridge University Press",
  year      =  2010,
  address   = "Cambridge, England"
}

@article{Esposito_2010, title={Entropy production as correlation between system and reservoir}, volume={12}, ISSN={1367-2630}, url={http://dx.doi.org/10.1088/1367-2630/12/1/013013}, DOI={10.1088/1367-2630/12/1/013013}, number={1}, journal={New Journal of Physics}, publisher={IOP Publishing}, author={Esposito, Massimiliano and Lindenberg, Katja and Van den Broeck, Christian}, year={2010}, month=jan, pages={013013} }

@article{Tanimura_exact_2020,
    author = {Tanimura, Yoshitaka},
    title = {Numerically “exact” approach to open quantum dynamics: The hierarchical equations of motion (HEOM)},
    journal = {The Journal of Chemical Physics},
    volume = {153},
    number = {2},
    pages = {020901},
    year = {2020},
    month = {07},
    issn = {0021-9606},
    doi = {10.1063/5.0011599},
    url = {https://doi.org/10.1063/5.0011599},
}

@article{efficient_heom,
    author = {Shi, Qiang and Chen, Liping and Nan, Guangjun and Xu, Rui-Xue and Yan, YiJing},
    title = {Efficient hierarchical Liouville space propagator to quantum dissipative dynamics},
    journal = {The Journal of Chemical Physics},
    volume = {130},
    number = {8},
    pages = {084105},
    year = {2009},
    month = {02},
    issn = {0021-9606},
    doi = {10.1063/1.3077918},
    url = {https://doi.org/10.1063/1.3077918},
}

@article{tanimura_heat_current,
    author = {Kato, Akihito and Tanimura, Yoshitaka},
    title = {Quantum heat current under non-perturbative and non-Markovian conditions: Applications to heat machines},
    journal = {The Journal of Chemical Physics},
    volume = {145},
    number = {22},
    pages = {224105},
    year = {2016},
    month = {12},
    issn = {0021-9606},
    doi = {10.1063/1.4971370},
    url = {https://doi.org/10.1063/1.4971370},
}

@article{tanimura_two_qubit_paper,
    author = {Kato, Akihito and Tanimura, Yoshitaka},
    title = {Quantum heat transport of a two-qubit system: Interplay between system-bath coherence and qubit-qubit coherence},
    journal = {The Journal of Chemical Physics},
    volume = {143},
    number = {6},
    pages = {064107},
    year = {2015},
    month = {08},
    issn = {0021-9606},
    doi = {10.1063/1.4928192},
    url = {https://doi.org/10.1063/1.4928192},
}

@article{Landi_ness_paper,
  title = {Nonequilibrium boundary-driven quantum systems: Models, methods, and properties},
  author = {Landi, Gabriel T. and Poletti, Dario and Schaller, Gernot},
  journal = {Rev. Mod. Phys.},
  volume = {94},
  issue = {4},
  pages = {045006},
  numpages = {58},
  year = {2022},
  month = {Dec},
  publisher = {American Physical Society},
  doi = {10.1103/RevModPhys.94.045006},
  url = {https://link.aps.org/doi/10.1103/RevModPhys.94.045006}
}

@article{Hayden_2022, title={A canonical Hamiltonian for open quantum systems}, volume={55}, ISSN={1751-8121}, url={http://dx.doi.org/10.1088/1751-8121/ac65c2}, DOI={10.1088/1751-8121/ac65c2}, number={22}, journal={Journal of Physics A: Mathematical and Theoretical}, publisher={IOP Publishing}, author={Hayden, Patrick and Sorce, Jonathan}, year={2022}, month=may, pages={225302} }

@article{Colla_canHam_paper,
  title = {Open-system approach to nonequilibrium quantum thermodynamics at arbitrary coupling},
  author = {Colla, Alessandra and Breuer, Heinz-Peter},
  journal = {Phys. Rev. A},
  volume = {105},
  issue = {5},
  pages = {052216},
  numpages = {8},
  year = {2022},
  month = {May},
  publisher = {American Physical Society},
  doi = {10.1103/PhysRevA.105.052216},
  url = {https://link.aps.org/doi/10.1103/PhysRevA.105.052216}
}

@misc{tiwari2025_finite_bath,
      title={A Comprehensive Approach to Finite-Bath Open Quantum Systems: Exact Dynamics}, 
      author={Devvrat Tiwari and Subhashish Banerjee},
      year={2025},
      eprint={2511.07180},
      archivePrefix={arXiv},
      primaryClass={quant-ph},
      url={https://arxiv.org/abs/2511.07180}, 
}

@Article{Colla2025,
author={Colla, Alessandra
and Hasse, Florian
and Palani, Deviprasath
and Schaetz, Tobias
and Breuer, Heinz-Peter
and Warring, Ulrich},
title={Observing time-dependent energy level renormalisation in an ultrastrongly coupled open system},
journal={Nature Communications},
year={2025},
month={Mar},
day={13},
volume={16},
number={1},
pages={2502},
issn={2041-1723},
doi={10.1038/s41467-025-57840-4},
url={https://doi.org/10.1038/s41467-025-57840-4}
}

@article{Maekawa_2013, title={Spin Current: Experimental and Theoretical Aspects}, volume={82}, ISSN={1347-4073}, url={http://dx.doi.org/10.7566/jpsj.82.102002}, DOI={10.7566/jpsj.82.102002}, number={10}, journal={Journal of the Physical Society of Japan}, publisher={Physical Society of Japan}, author={Maekawa, Sadamichi and Adachi, Hiroto and Uchida, Ken-ichi and Ieda, Jun’ichi and Saitoh, Eiji}, year={2013}, month=oct, pages={102002} }

@article{Bloch_2016,
  title = {Spin Pumping and Measurement of Spin Currents in Optical Superlattices},
  author = {Schweizer, C. and Lohse, M. and Citro, R. and Bloch, I.},
  journal = {Phys. Rev. Lett.},
  volume = {117},
  issue = {17},
  pages = {170405},
  numpages = {5},
  year = {2016},
  month = {Oct},
  publisher = {American Physical Society},
  doi = {10.1103/PhysRevLett.117.170405},
  url = {https://link.aps.org/doi/10.1103/PhysRevLett.117.170405}
}

@article{Poletti_2018,
  title = {Perfect Diode in Quantum Spin Chains},
  author = {Balachandran, Vinitha and Benenti, Giuliano and Pereira, Emmanuel and Casati, Giulio and Poletti, Dario},
  journal = {Phys. Rev. Lett.},
  volume = {120},
  issue = {20},
  pages = {200603},
  numpages = {5},
  year = {2018},
  month = {May},
  publisher = {American Physical Society},
  doi = {10.1103/PhysRevLett.120.200603},
  url = {https://link.aps.org/doi/10.1103/PhysRevLett.120.200603}
}

@article{Poulsen_2022,
  title = {Entanglement-enhanced quantum rectification},
  author = {Poulsen, Kasper and Santos, Alan C. and Kristensen, Lasse B. and Zinner, Nikolaj T.},
  journal = {Phys. Rev. A},
  volume = {105},
  issue = {5},
  pages = {052605},
  numpages = {12},
  year = {2022},
  month = {May},
  publisher = {American Physical Society},
  doi = {10.1103/PhysRevA.105.052605},
  url = {https://link.aps.org/doi/10.1103/PhysRevA.105.052605}
}

@article{Poulsen_2024,
  title = {Heat-based circuits using quantum rectification},
  author = {Poulsen, Kasper and Zinner, Nikolaj T.},
  journal = {Phys. Rev. A},
  volume = {109},
  issue = {5},
  pages = {052223},
  numpages = {10},
  year = {2024},
  month = {May},
  publisher = {American Physical Society},
  doi = {10.1103/PhysRevA.109.052223},
  url = {https://link.aps.org/doi/10.1103/PhysRevA.109.052223}
}

@article{Stephen_clark_2024,
  title = {Giant Rectification in Strongly Interacting Driven Tilted Systems},
  author = {Mendoza-Arenas, Juan Jos\'e and Clark, Stephen R.},
  journal = {PRX Quantum},
  volume = {5},
  issue = {1},
  pages = {010341},
  numpages = {20},
  year = {2024},
  month = {Mar},
  publisher = {American Physical Society},
  doi = {10.1103/PRXQuantum.5.010341},
  url = {https://link.aps.org/doi/10.1103/PRXQuantum.5.010341}
}

@article{tiwari_strong,
    author = {Tiwari, Devvrat and Bose, Baibhab and Banerjee, Subhashish},
    title = {Strong coupling non-Markovian quantum thermodynamics of a finite-bath system},
    journal = {The Journal of Chemical Physics},
    volume = {162},
    number = {11},
    pages = {114104},
    year = {2025},
    month = {03},
    issn = {0021-9606},
    doi = {10.1063/5.0254029},
    url = {https://doi.org/10.1063/5.0254029}
}

@book{sbbook,
  address = {152 Beach Rd, 21-01 Gateway East, Singapore 189721},
  author = {Banerjee, Subhasish},
  publisher = {Springer Singapore},
  title = {Open Quantum Systems},
  year = 2018
}

@book{weiss,
  author    = {U Weiss}, 
  title     = {Quantum Dissipative Systems},
  publisher = {World Scientific},
  year      = 1999,
  volume    = 10,
  series    = {Series in Modern Condensed Matter Physics},
  edition   = 2,
  isbn      = {9789814494212}
}

@book{Louisell,
  title={Quantum Statistical Properties of Radiation},
  author={Louisell, W.H.},
  isbn={9780471547853},
  lccn={73000547},
  series={A Wiley-Interscience publication},
  url={https://books.google.co.in/books?id=NRlRAAAAMAAJ},
  year={1973},
  publisher={Wiley}
}

@article{GrabertSchrammIngold,
title = {Quantum Brownian motion: The functional integral approach},
journal = {Physics Reports},
volume = {168},
number = {3},
pages = {115-207},
year = {1988},
issn = {0370-1573},
doi = {https://doi.org/10.1016/0370-1573(88)90023-3},
url = {https://www.sciencedirect.com/science/article/pii/0370157388900233},
author = {Hermann Grabert and Peter Schramm and Gert-Ludwig Ingold}
}

@article{sbsterngerlach,
  title = {Quantum theory of a Stern-Gerlach system in contact with a linearly dissipative environment},
  author = {Banerjee, Subhashish and Ghosh, R.},
  journal = {Phys. Rev. A},
  volume = {62},
  issue = {4},
  pages = {042105},
  numpages = {8},
  year = {2000},
  month = {Sep},
  publisher = {American Physical Society},
  doi = {10.1103/PhysRevA.62.042105},
  url = {https://link.aps.org/doi/10.1103/PhysRevA.62.042105}
}

@article{sbqbm,
  title = {General quantum Brownian motion with initially correlated and nonlinearly coupled environment},
  author = {Banerjee, Subhashish and Ghosh, R.},
  journal = {Phys. Rev. E},
  volume = {67},
  issue = {5},
  pages = {056120},
  numpages = {13},
  year = {2003},
  month = {May},
  publisher = {American Physical Society},
  doi = {10.1103/PhysRevE.67.056120},
  url = {https://link.aps.org/doi/10.1103/PhysRevE.67.056120}
}

@article{sbdixit_2019,
  title = {Study of coherence and mixedness in meson and neutrino systems},
  volume = {79},
  ISSN = {1434-6052},
  url = {http://dx.doi.org/10.1140/epjc/s10052-019-6609-7},
  DOI = {10.1140/epjc/s10052-019-6609-7},
  number = {2},
  journal = {The European Physical Journal C},
  publisher = {Springer Science and Business Media LLC},
  author = {Dixit,  Khushboo and Naikoo,  Javid and Banerjee,  Subhashish and Alok,  Ashutosh Kumar},
  year = {2019},
  month = jan 
}

@article{sbthomas,
  title = {Thermodynamics of non-Markovian reservoirs and heat engines},
  author = {Thomas, George and Siddharth, Nana and Banerjee, Subhashish and Ghosh, Sibasish},
  journal = {Phys. Rev. E},
  volume = {97},
  issue = {6},
  pages = {062108},
  numpages = {8},
  year = {2018},
  month = {Jun},
  publisher = {American Physical Society},
  doi = {10.1103/PhysRevE.97.062108},
  url = {https://link.aps.org/doi/10.1103/PhysRevE.97.062108}
}

@article{gksl,
author = {Gorini, Vittorio and Kossakowski, Andrzej and Sudarshan, E. C. G.},
title = {Completely positive dynamical semigroups of N‐level systems},
journal = {Journal of Mathematical Physics},
volume = {17},
number = {5},
pages = {821-825},
year = {1976},
doi = {10.1063/1.522979},
URL = {https://aip.scitation.org/doi/abs/10.1063/1.522979}
}

@Article{lindblad,
author={Lindblad, G.},
title={On the generators of quantum dynamical semigroups},
journal={Communications in Mathematical Physics},
year={1976},
month={Jun},
day={01},
volume={48},
number={2},
pages={119-130},
issn={1432-0916},
doi={10.1007/BF01608499},
url={https://doi.org/10.1007/BF01608499}
}

@article{hcla_non_Markovianity,
  title = {Canonical form of master equations and characterization of non-Markovianity},
  author = {Hall, Michael J. W. and Cresser, James D. and Li, Li and Andersson, Erika},
  journal = {Phys. Rev. A},
  volume = {89},
  issue = {4},
  pages = {042120},
  numpages = {11},
  year = {2014},
  month = {Apr},
  publisher = {American Physical Society},
  doi = {10.1103/PhysRevA.89.042120},
  url = {https://link.aps.org/doi/10.1103/PhysRevA.89.042120}
}

@article{colloquium_NM_breuer,
  title = {Colloquium: Non-Markovian dynamics in open quantum systems},
  author = {Breuer, Heinz-Peter and Laine, Elsi-Mari and Piilo, Jyrki and Vacchini, Bassano},
  journal = {Rev. Mod. Phys.},
  volume = {88},
  issue = {2},
  pages = {021002},
  numpages = {24},
  year = {2016},
  month = {Apr},
  publisher = {American Physical Society},
  doi = {10.1103/RevModPhys.88.021002},
  url = {https://link.aps.org/doi/10.1103/RevModPhys.88.021002}
}

@article{Kumar_2018,
author = {Kumar, N. Pradeep and Banerjee, Subhashish and Srikanth, R. and Jagadish, Vinayak and Petruccione, Francesco},
title = {Non-Markovian Evolution: a Quantum Walk Perspective},
journal = {Open Systems \& Information Dynamics},
volume = {25},
number = {03},
pages = {1850014},
year = {2018},
doi = {10.1142/S1230161218500142},
URL = {https://doi.org/10.1142/S1230161218500142},
eprint = {https://doi.org/10.1142/S1230161218500142}
}

@Article{Sun2025,
author={Sun, Ke
and Kang, Mingyu
and Nuomin, Hanggai
and Schwartz, George
and Beratan, David N.
and Brown, Kenneth R.
and Kim, Jungsang},
title={Quantum simulation of spin-boson models with structured bath},
journal={Nature Communications},
year={2025},
month={Apr},
day={30},
volume={16},
number={1},
pages={4042},
issn={2041-1723},
doi={10.1038/s41467-025-59296-y},
url={https://doi.org/10.1038/s41467-025-59296-y}
}

@article{Mostame_2012,
doi = {10.1088/1367-2630/14/10/105013},
url = {https://doi.org/10.1088/1367-2630/14/10/105013},
year = {2012},
month = {oct},
publisher = {IOP Publishing},
volume = {14},
number = {10},
pages = {105013},
author = {Mostame, Sarah and Rebentrost, Patrick and Eisfeld, Alexander and Kerman, Andrew J and Tsomokos, Dimitris I and Aspuru-Guzik, Alán},
title = {Quantum simulator of an open quantum system using superconducting qubits: exciton transport in photosynthetic complexes},
journal = {New Journal of Physics},
}

@article{spin_boson_expt2,
  title = {Quantum simulation of the spin-boson model with a microwave circuit},
  author = {Lepp\"akangas, Juha and Braum\"uller, Jochen and Hauck, Melanie and Reiner, Jan-Michael and Schwenk, Iris and Zanker, Sebastian and Fritz, Lukas and Ustinov, Alexey V. and Weides, Martin and Marthaler, Michael},
  journal = {Phys. Rev. A},
  volume = {97},
  issue = {5},
  pages = {052321},
  numpages = {19},
  year = {2018},
  month = {May},
  publisher = {American Physical Society},
  doi = {10.1103/PhysRevA.97.052321},
  url = {https://link.aps.org/doi/10.1103/PhysRevA.97.052321}
}

@Article{Magazzu2018,
author={Magazz{\`u}, L.
and Forn-D{\'i}az, P.
and Belyansky, R.
and Orgiazzi, J.-L.
and Yurtalan, M. A.
and Otto, M. R.
and Lupascu, A.
and Wilson, C. M.
and Grifoni, M.},
title={Probing the strongly driven spin-boson model in a superconducting quantum circuit},
journal={Nature Communications},
year={2018},
month={Apr},
day={11},
volume={9},
number={1},
pages={1403},
issn={2041-1723},
doi={10.1038/s41467-018-03626-w},
url={https://doi.org/10.1038/s41467-018-03626-w}
}

@article{THORWART2004,
title = {Dynamics of the spin-boson model with a structured environment},
journal = {Chemical Physics},
volume = {296},
number = {2},
pages = {333-344},
year = {2004},
note = {The Spin-Boson Problem: From Electron Transfer to Quantum Computing ... to the 60th Birthday of Professor Ulrich Weiss},
issn = {0301-0104},
doi = {https://doi.org/10.1016/j.chemphys.2003.10.007},
url = {https://www.sciencedirect.com/science/article/pii/S0301010403005469},
author = {M Thorwart and E Paladino and M Grifoni}
}

@article{mahima2025,
author = {Yadav, Mahima and Tiwari, Devvrat and Banerjee, Subhashish},
title = {(Thermo-)Dynamics of the Spin-Boson Model in the Weak Coupling Regime: Application as a Quantum Battery},
journal = {Advanced Quantum Technologies},
volume = {n/a},
number = {n/a},
pages = {e00333},
doi = {https://doi.org/10.1002/qute.202500333},
url = {https://advanced.onlinelibrary.wiley.com/doi/abs/10.1002/qute.202500333},
year = {2025}
}

@article{Goan_2010,
  title = {Non-Markovian finite-temperature two-time correlation functions of system operators of a pure-dephasing model},
  author = {Goan, Hsi-Sheng and Jian, Chung-Chin and Chen, Po-Wen},
  journal = {Phys. Rev. A},
  volume = {82},
  issue = {1},
  pages = {012111},
  numpages = {9},
  year = {2010},
  month = {Jul},
  publisher = {American Physical Society},
  doi = {10.1103/PhysRevA.82.012111},
  url = {https://link.aps.org/doi/10.1103/PhysRevA.82.012111}
}

@ARTICLE{Goan_2011,
  title     = "{Non-Markovian} finite-temperature two-time correlation functions of system operators: beyond the quantum regression theorem",
  author    = "Goan, Hsi-Sheng and Chen, Po-Wen and Jian, Chung-Chin",
  journal   = "J. Chem. Phys.",
  publisher = "AIP Publishing",
  volume    =  134,
  number    =  12,
  pages     = "124112",
  month     =  mar,
  year      =  2011,
  language  = "en"
}

@article{Haase2018,
  title = {Fundamental limits to frequency estimation: a comprehensive microscopic perspective},
  volume = {20},
  ISSN = {1367-2630},
  url = {http://dx.doi.org/10.1088/1367-2630/aab67f},
  DOI = {10.1088/1367-2630/aab67f},
  number = {5},
  journal = {New Journal of Physics},
  publisher = {IOP Publishing},
  author = {Haase,  J F and Smirne,  A and Kołodyński,  J and Demkowicz-Dobrzański,  R and Huelga,  S F},
  year = {2018},
  month = may,
  pages = {053009}
}

@article{Dvira_rcm,
    author = {Anto-Sztrikacs, Nicholas and Ivander, Felix and Segal, Dvira},
    title = {Quantum thermal transport beyond second order with the reaction coordinate mapping},
    journal = {The Journal of Chemical Physics},
    volume = {156},
    number = {21},
    pages = {214107},
    year = {2022},
    month = {06},
    issn = {0021-9606},
    doi = {10.1063/5.0091133},
    url = {https://doi.org/10.1063/5.0091133},
}

@article{Shubrook_2025,
doi = {10.1088/2058-9565/adc6b7},
url = {https://doi.org/10.1088/2058-9565/adc6b7},
year = {2025},
month = {apr},
publisher = {IOP Publishing},
volume = {10},
number = {2},
pages = {025063},
author = {Shubrook, Mike and Iles-Smith, Jake and Nazir, Ahsan},
title = {Non-Markovian quantum heat statistics with the reaction coordinate mapping},
journal = {Quantum Science and Technology}
}

@article{tanimura-kubo,
author = {Tanimura ,Yoshitaka and Kubo ,Ryogo},
title = {Time Evolution of a Quantum System in Contact with a Nearly Gaussian-Markoffian Noise Bath},
journal = {Journal of the Physical Society of Japan},
volume = {58},
number = {1},
pages = {101-114},
year = {1989},
doi = {10.1143/JPSJ.58.101},
URL = {https://doi.org/10.1143/JPSJ.58.101}
}

@article{Amartya_bose_2023,
    author = {Bose, Amartya},
    title = {QuantumDynamics.jl: A modular approach to simulations of dynamics of open quantum systems},
    journal = {The Journal of Chemical Physics},
    volume = {158},
    number = {20},
    pages = {204113},
    year = {2023},
    month = {05},
    issn = {0021-9606},
    doi = {10.1063/5.0151483},
    url = {https://doi.org/10.1063/5.0151483}
}

@article{Garraway_pseudomode1,
  title = {Nonperturbative decay of an atomic system in a cavity},
  author = {Garraway, B. M.},
  journal = {Phys. Rev. A},
  volume = {55},
  issue = {3},
  pages = {2290--2303},
  numpages = {0},
  year = {1997},
  month = {Mar},
  publisher = {American Physical Society},
  doi = {10.1103/PhysRevA.55.2290},
  url = {https://link.aps.org/doi/10.1103/PhysRevA.55.2290}
}

@article{Linden_pseudomode2,
  title = {Nonequilibrium Dynamical Mean-Field Theory: An Auxiliary Quantum Master Equation Approach},
  author = {Arrigoni, Enrico and Knap, Michael and von der Linden, Wolfgang},
  journal = {Phys. Rev. Lett.},
  volume = {110},
  issue = {8},
  pages = {086403},
  numpages = {6},
  year = {2013},
  month = {Feb},
  publisher = {American Physical Society},
  doi = {10.1103/PhysRevLett.110.086403},
  url = {https://link.aps.org/doi/10.1103/PhysRevLett.110.086403}
}

@article{Plenio_pseudomode3,
  title = {Nonperturbative Treatment of non-Markovian Dynamics of Open Quantum Systems},
  author = {Tamascelli, D. and Smirne, A. and Huelga, S. F. and Plenio, M. B.},
  journal = {Phys. Rev. Lett.},
  volume = {120},
  issue = {3},
  pages = {030402},
  numpages = {6},
  year = {2018},
  month = {Jan},
  publisher = {American Physical Society},
  doi = {10.1103/PhysRevLett.120.030402},
  url = {https://link.aps.org/doi/10.1103/PhysRevLett.120.030402}
}

@article{Sassetti_1996,
  title = {Exact master equations for driven dissipative tight-binding models},
  author = {Grifoni, Milena and Sassetti, Maura and Weiss, Ulrich},
  journal = {Phys. Rev. E},
  volume = {53},
  issue = {3},
  pages = {R2033--R2036},
  numpages = {0},
  year = {1996},
  month = {Mar},
  publisher = {American Physical Society},
  doi = {10.1103/PhysRevE.53.R2033},
  url = {https://link.aps.org/doi/10.1103/PhysRevE.53.R2033}
}

@article{Sassetti_1990,
  title = {Universality in the dissipative two-state system},
  author = {Sassetti, Maura and Weiss, Ulrich},
  journal = {Phys. Rev. Lett.},
  volume = {65},
  issue = {18},
  pages = {2262--2265},
  numpages = {0},
  year = {1990},
  month = {Oct},
  publisher = {American Physical Society},
  doi = {10.1103/PhysRevLett.65.2262},
  url = {https://link.aps.org/doi/10.1103/PhysRevLett.65.2262}
}

@Article{kosloff,
AUTHOR = {Kosloff, Ronnie},
TITLE = {Quantum Thermodynamics: A Dynamical Viewpoint},
JOURNAL = {Entropy},
VOLUME = {15},
YEAR = {2013},
NUMBER = {6},
PAGES = {2100--2128},
URL = {https://www.mdpi.com/1099-4300/15/6/2100},
ISSN = {1099-4300},
DOI = {10.3390/e15062100}
}

@book{Gemmer2009,
  title = {Quantum Thermodynamics: Emergence of Thermodynamic Behavior Within Composite Quantum Systems},
  ISBN = {9783540705109},
  ISSN = {1616-6361},
  url = {http://dx.doi.org/10.1007/978-3-540-70510-9},
  DOI = {10.1007/978-3-540-70510-9},
  journal = {Lecture Notes in Physics},
  publisher = {Springer Berlin Heidelberg},
  author = {Gemmer,  Jochen and Michel,  M. and Mahler,  G\"{u}nter},
  year = {2009}
}

@article{Rivas_strong_coupling,
  title = {Strong Coupling Thermodynamics of Open Quantum Systems},
  author = {Rivas, \'Angel},
  journal = {Phys. Rev. Lett.},
  volume = {124},
  issue = {16},
  pages = {160601},
  numpages = {7},
  year = {2020},
  month = {Apr},
  publisher = {American Physical Society},
  doi = {10.1103/PhysRevLett.124.160601},
  url = {https://link.aps.org/doi/10.1103/PhysRevLett.124.160601}
}

@book{deffner_book,
author = {Deffner, Sebastian and Campbell, Steve},
title = {Quantum Thermodynamics},
publisher = {Morgan \& Claypool Publishers},
year = {2019},
series = {2053-2571},
isbn = {978-1-64327-658-8},
url = {https://dx.doi.org/10.1088/2053-2571/ab21c6},
doi = {10.1088/2053-2571/ab21c6}
}

@article{Segal_spin_boson_rectifier,
  title = {Spin-Boson Thermal Rectifier},
  author = {Segal, Dvira and Nitzan, Abraham},
  journal = {Phys. Rev. Lett.},
  volume = {94},
  issue = {3},
  pages = {034301},
  numpages = {4},
  year = {2005},
  month = {Jan},
  publisher = {American Physical Society},
  doi = {10.1103/PhysRevLett.94.034301},
  url = {https://link.aps.org/doi/10.1103/PhysRevLett.94.034301}
}

@article{diode1,
  title = {Thermal Diode: Rectification of Heat Flux},
  author = {Li, Baowen and Wang, Lei and Casati, Giulio},
  journal = {Phys. Rev. Lett.},
  volume = {93},
  issue = {18},
  pages = {184301},
  numpages = {4},
  year = {2004},
  month = {Oct},
  publisher = {American Physical Society},
  doi = {10.1103/PhysRevLett.93.184301},
  url = {https://link.aps.org/doi/10.1103/PhysRevLett.93.184301}
}

@article{thermal_diode,
  title = {Quantum thermal diode based on two interacting spinlike systems under different excitations},
  author = {Ordonez-Miranda, Jose and Ezzahri, Youn\`es and Joulain, Karl},
  journal = {Phys. Rev. E},
  volume = {95},
  issue = {2},
  pages = {022128},
  numpages = {7},
  year = {2017},
  month = {Feb},
  publisher = {American Physical Society},
  doi = {10.1103/PhysRevE.95.022128},
  url = {https://link.aps.org/doi/10.1103/PhysRevE.95.022128}
}

@article{transistor,
  title = {Quantum Thermal Transistor},
  author = {Joulain, Karl and Drevillon, J\'er\'emie and Ezzahri, Youn\`es and Ordonez-Miranda, Jose},
  journal = {Phys. Rev. Lett.},
  volume = {116},
  issue = {20},
  pages = {200601},
  numpages = {5},
  year = {2016},
  month = {May},
  publisher = {American Physical Society},
  doi = {10.1103/PhysRevLett.116.200601},
  url = {https://link.aps.org/doi/10.1103/PhysRevLett.116.200601}
}

@article{QWB,
  title = {Quantum Wheatstone Bridge},
  author = {Poulsen, Kasper and Santos, Alan C. and Zinner, Nikolaj T.},
  journal = {Phys. Rev. Lett.},
  volume = {128},
  issue = {24},
  pages = {240401},
  numpages = {6},
  year = {2022},
  month = {Jun},
  publisher = {American Physical Society},
  doi = {10.1103/PhysRevLett.128.240401},
  url = {https://link.aps.org/doi/10.1103/PhysRevLett.128.240401}
}

@article{Baowen_review,
  title = {Colloquium: Phononics: Manipulating heat flow with electronic analogs and beyond},
  author = {Li, Nianbei and Ren, Jie and Wang, Lei and Zhang, Gang and H\"anggi, Peter and Li, Baowen},
  journal = {Rev. Mod. Phys.},
  volume = {84},
  issue = {3},
  pages = {1045--1066},
  numpages = {0},
  year = {2012},
  month = {Jul},
  publisher = {American Physical Society},
  doi = {10.1103/RevModPhys.84.1045},
  url = {https://link.aps.org/doi/10.1103/RevModPhys.84.1045}
}

@article{Dubi_review,
  title = {Colloquium: Heat flow and thermoelectricity in atomic and molecular junctions},
  author = {Dubi, Yonatan and Di Ventra, Massimiliano},
  journal = {Rev. Mod. Phys.},
  volume = {83},
  issue = {1},
  pages = {131--155},
  numpages = {0},
  year = {2011},
  month = {Mar},
  publisher = {American Physical Society},
  doi = {10.1103/RevModPhys.83.131},
  url = {https://link.aps.org/doi/10.1103/RevModPhys.83.131}
}

@article{BENENTI2017,
title = {Fundamental aspects of steady-state conversion of heat to work at the nanoscale},
journal = {Physics Reports},
volume = {694},
pages = {1-124},
year = {2017},
issn = {0370-1573},
doi = {https://doi.org/10.1016/j.physrep.2017.05.008},
url = {https://www.sciencedirect.com/science/article/pii/S0370157317301540},
author = {Giuliano Benenti and Giulio Casati and Keiji Saito and Robert S. Whitney}
}

@article{Majland_2020,
  title = {Quantum thermal transistor in superconducting circuits},
  author = {Majland, Marco and Christensen, Kasper Sangild and Zinner, Nikolaj Thomas},
  journal = {Phys. Rev. B},
  volume = {101},
  issue = {18},
  pages = {184510},
  numpages = {10},
  year = {2020},
  month = {May},
  publisher = {American Physical Society},
  doi = {10.1103/PhysRevB.101.184510},
  url = {https://link.aps.org/doi/10.1103/PhysRevB.101.184510}
}

@Article{Cerisola_2025,
author={Fedele, Federico
and Cerisola, Federico
and Bresque, Lea
and Vigneau, Florian
and Monsel, Juliette
and Tabanera-Bravo, Jorge
and Aggarwal, Kushagra
and Dexter, Jonathan
and Sevitz, Sofia
and Dunlop, Joe
and Auffeves, Alexia
and Parrondo, Juan MR
and Palyi, Andras
and Anders, Janet
and Ares, Natalia},
title={Coupling a single spin to the motion of a carbon nanotube},
journal={Nature Communications},
year={2025},
month={Dec},
day={13},
volume={16},
number={1},
pages={11454},
issn={2041-1723},
doi={10.1038/s41467-025-66331-5},
url={https://doi.org/10.1038/s41467-025-66331-5}
}

\end{document}